\documentstyle[twocolumn,aps]{revtex}%[preprint,fleqn,eqsecnum]
\begin{document} \draft 
\title{ Berry phases and pairing symmetry
in Holstein-Hubbard polaron systems}
\author{ K. Yonemitsu } 
\address{ Department of Theoretical Studies, Institute for
 Molecular Science, Okazaki, Aichi 444-8585, Japan }
\author{ J. Zhong, and H.-B. Sch\"uttler }
\address{ Center for Simulational Physics, Department of Physics
 and Astronomy, University of Georgia, Athens, Georgia 30602 }
\date{\today} \maketitle

\begin{abstract}
We study the tunneling dynamics of 
dopant-induced hole polarons which are 
self-localized by electron-phonon coupling in a 
two-dimensional antiferromagnet. Our treatment is based on
a path integral formulation of the adiabatic (Born-Oppenheimer)
approximation, combined with many-body tight-binding,
instanton, constrained lattice dynamics,
and many-body exact diagonalization techniques.
The applicability and limitations
of the adiabatic approximation in polaron tunneling problems
are discussed in detail and adiabatic results are compared
to exact numerical results for a two-site polaron problem.
Our results are mainly based on the Holstein-$ tJ $ 
and, for comparison, on the Holstein-Hubbard model.
We also study the effects of 2nd neighbor hopping and  
long-range electron-electron Coulomb repulsion.
The polaron tunneling dynamics is 
mapped onto an effective low-energy Hamiltonian 
which takes the form of a fermion tight-binding model with 
occupancy dependent, predominantly 2nd and 3rd neighbor 
tunneling matrix elements, excluded double occupancy, 
and an effective intersite charge interactions.
Antiferromagnetic spin correlations in the original 
many-electron Hamiltonian are reflected by an attractive
contribution to the 1st neighbor charge interaction and 
by Berry phase factors which determine the signs of effective 
polaron tunneling matrix elements. In the two-polaron case, 
these phase factors lead to polaron pair
wave functions of either $d_{x^2-y^2}$-wave symmetry or $ p $-wave 
symmetry with zero and nonzero total pair momentum, respectively. 
Implications for the doping dependent
isotope effect, pseudo-gap and $T_c$ 
of a superconducting polaron pair condensate 
are discussed and compared to observed properties
of the cuprate high-$T_c$ materials.
\end{abstract} 

\section{INTRODUCTION}

The symmetry of the superconducting order parameter 
in the cuprate high-$T_c$ superconductors had been 
controversial\cite{Levi} before 
phase-sensitive experiments firmly established 
the $d_{x^2-y^2}$-wave pairing symmetry in YBa$_2$Cu$_3$O$_7$, 
using tricrystal ring magnetometry\cite{tricrystal}, 
SQUID interferometry\cite{SQUID}, and single-junction 
modulation\cite{junction}.
Migdal-Eliashberg-type diagrammatic theories find 
$d$-wave pairing to be favored by antiferromagnetic (AF) 
spin-fluctuation exchange\cite{Scalapino} and 
$s$-wave pairing by the conventional electron-phonon 
mechanism.\cite{Scalapino,Pao}
There is indeed strong experimental evidence for 
the importance of {\em both \/} AF spin correlations\cite{afm_exp} 
{\em and \/} electron-phonon interactions\cite{phonon_exp} 
in the cuprates.
However, when combined in the diagrammatic approach, 
the two mechanisms are mutually destructive, since 
$d$-wave pairing is strongly suppressed 
by phonons and $s$-wave pairing is suppressed
by AF spin fluctuations, respectively.
Also, the magnitude of the observed isotope effect in cuprate systems 
away from ``optimal'' doping\cite{isotope_exp} points towards 
an unusually strong electron-phonon effect which cannot be 
accounted for in the diagrammatic approaches.\cite{Pao}

Strong-coupling studies,\cite{YBL,Zhong,Roeder,recent_EP} 
going beyond the Migdal-Eliashberg regime, suggest that 
the AF spin correlations themselves can effectively 
enhance the electron-phonon effect, by lowering 
the electron-phonon coupling threshold for polaron formation, 
that is, the threshold for electron-phonon induced 
self-localization\cite{Holstein} of the dopant-induced carriers 
in the CuO$_2$ planes.
In the present paper, we show how the tunneling dynamics 
of such self-localized holes in an AF correlated spin background 
may lead to $d$- and other non-$s$-wave pairing states which
are {\it not}  suppressed by coupling 
to the lattice degrees of freedom.

A Berry phase factor in finite systems with time-reversal symmetry 
has been relevant to the observation of half-odd-integer quantum 
numbers in the spectrum of the Na$_3$ molecule,\cite{Delacretaz}, 
to the cross section of the H+H$_2$ reaction and its isotope 
analogs,\cite{Wu} and to the problem of integer vs. half-odd-integer 
spin tunneling in anisotropic potentials.\cite{Loss} 
Contributions to the pair binding energy in the C$_{60}$ molecule 
have also been discussed in terms of Berry phase 
arguments.\cite{Auerbach} 
In the present case, the non-$s$-wave symmetry is caused 
by a ($-1$) Berry phase factor, associated with predominantly 
second- and third-neighbor polaron tunneling processes.
It also determines the total momentum: 
the one-polaron ground state has a momentum on the Fermi surface 
of the half-filled tight-binding model on the square lattice.
The dynamics of few hole polarons reflects 
the local AF spin correlations of many electrons 
through the Berry phase factor.

This paper is organized as follows: 
In Sec.~\ref{sec:model}, we introduce the basic
Holstein-Hubbard and Holstein-$ tJ $ model
Hamiltonians, and their extensions to include
2nd neighbor hopping or long-range Coulomb repulsion.
We then derive the effective action for the lattice 
degrees of freedom in the adiabatic approximation.
In Sec.~\ref{sec:two-site}, we illustrate the basic physical
principles and formal concepts
of our adiabatic treatment of the polaron tunneling
in the context of a simple two-site model.
In Sec.~\ref{sec:validity}, we discuss the conditions
under which the adiabatic approximation is valid,
as well as its limitations when applied to polaronic
systems on large / macroscopic lattice systems.
In particular, we clear up some recent misunderstandings
concerning the applicability of the adiabatic approach to
polaronic systems.
In Sec.~\ref{sec:instanton}, we use an instanton approach
to elucidate the basic structure of the
low-energy tunneling dynamics of hole polarons in
Holstein-Hubbard or Holstein-$ tJ $ systems near half-filling.
We show that the dynamics of such hole polarons 
is governed by an effective tight-binding Hamiltonian 
which includes 2nd and 3rd neighbor hopping matrix elements
and a 1st neighbor attraction.
In Sec.~\ref{sec:symmetry}, we discuss the Berry phase factors
and, with the help of lattice symmetry operations, we
show how such phases can be properly assigned to
each segment of a closed tunneling path.
The Berry phase factors are then interpreted in
terms a quasiparticle statistics 
and internal symmetries of 
the many-electron wave functions.
In Sec.~\ref{sec:phase}, we analytically solve the effective model 
to show how the Berry phase factors determine the total momenta and 
internal symmetries of the few-hole-polaron wave functions.
In Sec.~\ref{sec:effective}, we report numerical results
for the effective polaron hopping and 
effective pair binding energy as functions of the phonon frequency
and electron-phonon coupling strength.
In Sec.~\ref{sec:p_liquid}, we discuss 
the implications
of our numerical results for a possible superconducting
pairing instability, the isotope effect and the
pseudo-gap in a hole polaron liquid at finite
doping concentration in the nearly half-filled 
Holstein-$ tJ $ and -Hubbard systems and
compare to experimental observations in the
cuprates.
In Sec.~\ref{sec:summary}, we summarize the present work.
Part of the results presented in this paper were reported briefly 
in an unpublished paper and proceedings.\cite{YZS}

\section{MODEL AND EFFECTIVE ACTION}
\label{sec:model}

We use mainly the Holstein-$ tJ $ model\cite{Zhong,Roeder} 
and occasionally the Holstein-Hubbard model for comparison.
Later, we also include 2nd neighbor electron hopping 
and/or long-range electron-electron repulsion terms in the model.
The total Hamiltonian is of the general form 
\begin{equation}
H = H_{\rm e} + H_{\rm e-ph} + H_{\rm ph}
\;,\label{eq:model}
\end{equation}
where $H_{\rm e}$ is the purely electronic $ tJ $ or Hubbard model
part, defined on a two-dimensional (2D) square lattice with lattice 
sites $j = 1 \dots N$ and on-site electron occupation numbers $n_j$,
as specified below.
\begin{equation}
H_{\rm e-ph}=C \sum_j u_j n_j 
\label{eq:h_elph}
\end{equation}
is the Holstein electron-phonon (EP) interaction, coupling 
the local oscillator displacement $u_j$ to the electron
on-site occupation $n_j$ with an EP coupling constant $C$ and 
\begin{equation}
H_{\rm ph}= \frac{K}{2} \sum_j u_j^2 
  + \frac{1}{2M} \sum_j p_j^2
\equiv H_{\rm K} + H_{\rm M}
\label{eq:h_phon}
\end{equation}
describes the non-interacting Einstein phonon system,
consisting of the bare harmonic lattice potential $H_{\rm K}$,
with restoring force constant $K$, and of the lattice kinetic
energy $H_{\rm M}$ with an atomic mass $M$ and
conjugate momenta $ p_j \equiv -i \hbar \partial / 
\partial u_j $. If we rescale to dimensionless
displacements and conjugate momenta
\begin{equation}
\bar{u}_j \equiv u_j / u_{\rm P},\;\;\;\;\;
\bar{p}_j\equiv -i \partial / \partial \bar{u}_j 
\;,
\label{eq:u_bar}
\end{equation}
with the small polaron shift
\begin{equation}
u_{\rm P} \equiv \frac{C}{K}
\;,
\label{eq:u_pol}
\end{equation}
then $H_{\rm e-ph}$ and $H_{\rm ph}$
can be completely parametrized in terms of only two
characteristic energies, the bare Einstein phonon energy
\begin{equation}
\Omega \equiv \hbar \Big(\frac{K}{M}\Big)^{\frac12}
\;,
\label{eq:omega}
\end{equation}
and the ionic-limit ($t\to0$)
small polaron binding energy
\begin{equation}
E_{\rm P}\equiv \frac{C^2} {K}
\;.
\label{eq:e_pol}
\end{equation}
All results in the following are therefore stated in terms
of $u_{\rm P}$, $\Omega$ and $E_{\rm P}$ only \cite{Zhong,Holstein}.

The $ tJ $ model is written as\cite{Anderson} 
\begin{equation}
H_{\rm e} = -t \sum_{\langle i,j \rangle, \sigma} \left( 
c_{i \sigma}^\dagger c_{j \sigma} + {\rm H.c.} \right)
+J \sum_{\langle i,j \rangle} \left( 
{\bf S}_i \cdot {\bf S}_j - \frac{n_i n_j}{4} \right)
\label{eq:h_tj}
\end{equation}
with 1st neighbor electron hopping $t$ 
and AF exchange coupling $J$. Here, $c_{i \sigma}$ annihilates 
an electron with spin $ \sigma $ at site $i$, 
$ n_{i \sigma} = c_{i \sigma}^\dagger c_{i \sigma} $, 
$ n_i = \sum_\sigma n_{i \sigma} $, 
$ {\bf S}_i = \frac12 \sum_{\alpha, \beta}
c_{i \alpha}^\dagger {\bf \sigma}_{\alpha \beta} c_{i \beta} $
with ${\bf \sigma}\equiv(\sigma_x,\sigma_y,\sigma_z)$ denoting the
vector of Pauli spin matrices.
The Hilbert space is restricted to states with no double
occupancy at any site $j$, i.e., $n_j=0,1$ only.

The Hubbard model is written as 
\begin{equation}
H_{\rm e} = -t \sum_{\langle i,j \rangle, \sigma} \left( 
c_{i \sigma}^\dagger c_{j \sigma} + {\rm H.c.} \right) 
+U \sum_i n_{i \uparrow} n_{i \downarrow} 
\label{eq:h_hub}
\end{equation}
with on-site repulsion $U$ and no restrictions on the
on-site occupancy, i.e., $n_j=0,1,2$.
In the following, we set $ \hbar \equiv 1 $, $ t \equiv 1 $ and use 
$ J = 0.5t $ or $ U = 8t $ in the $ tJ $ or Hubbard model, 
respectively, unless stated otherwise.

In addition to the standard $ tJ $ and Hubbard electronic model,
we will also study the effects of additional, potentially
important electronic terms, the 2nd neighbor hopping $H_{t'}$ 
and the long-range Coulomb repulsion $H_{\rm lc}$. Namely,
\begin{equation}
H_{t'}=
-t'\sum_{\{ i,j \}, \sigma} \left( 
c_{i \sigma}^\dagger c_{j \sigma} + {\rm H.c.} \right) 
\label{eq:h_t2}
\end{equation}
where $\{ i,j \}$ denotes 2nd neighbor bonds and $t'$ is the 
corresponding 2nd neighbor matrix element.
The long-range $1/r$ Coulomb repulsion is
\begin{equation}
H_{\rm lc}= \frac12
V_{\rm C} \sum_{i\neq j} \frac{ n_i n_j }{ | r_{ij} | } 
\label{eq:h_lc}
\end{equation}
where $i$ and $j$ are summed independently over all sites
excluding $ i = j $ and $r_{ij}$ denotes the vector pointing
from $i$ to $j$, measured in units of the 2D lattice constant
$a\equiv 1$. On a lattice with periodic boundary conditions
we make the definition of $| r_{ij}|$ unique by requiring
$r_{ij}$ to be a vector of the shortest possible length
connecting $i$ to $j$, subject to all possible periodic boundary 
shifts.
The matrix element $V_{\rm C}$ is thus the Coulomb repulsion
energy between two electrons at 1st neighbor distance.

To study the tunneling dynamics of self-localized holes, 
we consider the path integrals for transition amplitudes 
in imaginary time in the Born-Oppenheimer (adiabatic)
approximation. Following the standard Feynman-Trotter
approach \cite{Feynman-Hibbs}, we break up the Hamiltonian 
in the imaginary-time evolution operator
\begin{equation}
e^{-\beta H}= \lim_{L\to\infty}
\Big( e^{-\Delta\tau H_0} e^{-\Delta\tau H_{\rm M}}\Big)^L
\label{eq:e_bh}
\end{equation}
where $\Delta\tau\equiv\beta/L$, $H_{\rm M}$ is the lattice 
kinetic energy defined in Eq.~(\ref{eq:h_phon}) 
and the 0th order part $H_0\equiv H-H_{\rm M}$
commutes with all lattice displacement operators $u_j$.
At each time slice $\tau_k\equiv k\Delta\tau$, with $k=1\dots L$, 
we now insert a complete set of electron-phonon basis states 
$\mid\chi^{(\kappa)}_u\rangle$ 
which are chosen to be simultaneous eigenstates 
of $H_0$ and of all $u_j$. 
They can be written in the form
\begin{equation}
\mid\chi_u^{(\kappa)}\rangle = \mid\Psi^{(\kappa)}(u)\rangle 
                      \times\mid \Phi_u\rangle
\label{eq:chi_u}
\end{equation}
where $\mid\Phi_u\rangle$ is the lattice part 
and $\mid\Psi^{(\kappa)}(u)\rangle$
the electronic part of $\mid\chi_u^{(\kappa)}\rangle$.
Written in 1st quantized notation, the lattice part is simply
\begin{equation}
\Phi_u(x)=\delta(u-x) = \prod_j \delta(u_j-x_j)
\label{eq:phi_u}
\end{equation}
with lattice coordinate vectors
$x\equiv(x_1\dots x_N)$ and $u\equiv(u_1\dots u_N)$.
The electronic part $\mid\Psi^{(\kappa)}(u)\rangle$ 
denotes the $\kappa$-th electronic eigenstate 
of the 0th order adiabatic Hamiltonian
\begin{equation}
H_0(u)= H_{\rm e} + H_{\rm e-ph}(u) + H_{\rm K}(u)
\;,
\label{eq:h_0}
\end{equation}
at fixed $u$. That is, $H_0(u)$ is
defined to act only on the electronic degrees of freedom
at {\it fixed} ($c$-number) lattice displacement coordinates 
$u\equiv(u_1\dots u_N)$ and 
\begin{equation}
H_0(u)\mid\Psi^{(\kappa)}(u)\rangle= 
W_0^{(\kappa)}(u) \mid\Psi^{(\kappa)}(u)\rangle
\label{eq:w_0}
\end{equation}
where $\mid\Psi^{(\kappa)(u)}\rangle$ and its eigenenergy 
$W_0^{(\kappa)}(u)$ depend parametrically on the lattice 
displacements $u$.
The exact imaginary time evolution under $H$ can thus be represented
by a path integral with a Euclidean action, written at finite $L$ as
\begin{eqnarray}
S[u(\tau),\kappa(\tau)]&&= \sum_{k=1}^{L} \Big[
(M/2) \sum_j \frac{[ u_j(\tau_k) - u_j(\tau_{k-1}) ]^2}{\Delta \tau} 
\nonumber \\
&&
+ \Delta \tau W_0^{(\kappa_k)}(u(\tau_k)) 
\nonumber \\
&& 
- \ln \langle \Psi^{(\kappa_k)}(u(\tau_k)) \mid 
              \Psi^{(\kappa_{k-1})}(u(\tau_{k-1})) \rangle \Big]
\;.
\label{eq:s_exact}
\end{eqnarray}
The path integration is to be carried out both over the
continuous lattice coordinates 
$u(\tau_k)\equiv(u_1(\tau_k)\dots u_N(\tau_k)$)
and over the discrete electronic quantum numbers 
$\kappa_k\equiv\kappa(\tau_k)$.
%XXXX
%Notice that, due to the trace operation in the partition function,
%only closed paths, obeying periodic boundary conditions in
%imaginary time, enter into the path integrals above.
%Analogous expressions can be obtained for the effective action
%of open paths, arising in the more general problem
%of imaginary time transition amplitudes, the only difference
%to the closed-path case being the absence of the imaginary-time
%periodic boundary requirement.
%XXXX

In the 0th order adiabatic approximation, corresponding formally
to the $M\to\infty$ limit, one neglects the imaginary time
evolution of $u$ altogether and replaces $u_k$ by a $\tau$-independent
classical field. The 1st order adiabatic approximation
restores the $\tau$-dependence of the lattice coordinates $u$,
under the simplifying assumption that the electrons follow the
motion of the lattice adiabatically. That is, the path integration
is restricted to configurations where, during $\tau$-evolution,
the electrons remain in the same eigenstate, i.e., 
$\kappa_k=\kappa_{k-1}\equiv\kappa={\rm const}$. Transitions between
different electronic eigenstates $\kappa_k\neq\kappa_{k-1}$ are 
neglected.
Formally, this approximation restores the 
leading order $1/M$ corrections
to the lattice dynamics.
At sufficiently low temperatures, one restricts the path
integral further to include only the electronic groundstate
$\kappa=0$. Suppressing the $(\kappa)$-superscript altogether,
one then arrives at the standard 1st order adiabatic 
(Born-Oppenheimer) approximation, with an effective Euclidean action
\begin{eqnarray}
S_{\rm Ad} [u(\tau)] &&= \sum_{k=1}^{L} \big[
\frac{M}{2}\sum_j 
\frac{[ u_j(\tau_k) - u_j(\tau_{k-1}) ]^2} {\Delta \tau} 
\nonumber \\
&& + \Delta \tau W_0(u(\tau_k)) 
\nonumber \\
&&- \ln \langle \Psi(u(\tau_k)) \mid \Psi(u(\tau_{k-1})) \rangle \big]
\;.
\label{eq:s_eff0}
\end{eqnarray}
Note that $S_{\rm Ad}$ depends explicitly only on the $u$-coordinates
of the lattice.
The first ($M/2$-) term is the 
standard form of the lattice kinetic
energy for discretized imaginary time (finite $L$). 
The electronic groundstate 
energy $W_0(u)$ plays the role 
of a 0th-order (in $1/M$) effective lattice potential energy.
The last term, containing the logarithms of the electronic groundstate
wavefunction overlaps at adjacent time slices during $\tau$-evolution,
contains the Berry phase and $1/M$ corrections to the potential
energy, as we will now discuss.

In $ \exp( -S_{\rm Ad} [u(\tau)] ) $, the overlap product 
\begin{equation}
Q [u(\tau)] \equiv \prod_{k=1}^{L} 
\langle \Psi(u(\tau_k)) \mid \Psi(u(\tau_{k-1})) \rangle
\label{eq:overlap}
\end{equation}
enters which contains the Berry phase factor, 
\begin{equation}
\exp( -i \theta [u(\tau)] ) 
\equiv Q [u(\tau)] /\mid Q [u(\tau)] \mid 
\;,
\end{equation}
{\it i.e.},
$
\theta [u(\tau)] = -
{\rm Im}\ln( Q [u(\tau)]).
$
Due to time-reversal symmetry, all 
$\mid\Psi(u(\tau_k))\rangle$ have
real amplitudes in an appropriately chosen electron 
basis and hence the phase factor is real: 
$ \exp( -i \theta [u(\tau)] ) 
= {\rm sign}( Q [u(\tau)] ) $.
Taking $ L \rightarrow \infty $, we can also rewrite
Re$(\ln Q[u(\tau)]) \equiv \ln \mid Q[u(\tau)] \mid $ in 
$ S_{\rm Ad} [u(\tau)] $ 
as a $ 1/M $ correction to the effective lattice potential 
which thus becomes
\begin{equation}
W(u) \equiv W_0(u) + W_1(u) 
\;,
\end{equation}
with $W_1$ given by 
\begin{equation}
W_1(u) = \frac1{2M} \sum_j 
\langle \partial_{u_j} \Psi(u) \mid \partial_{u_j} \Psi(u) \rangle
\;.
\label{eq:w_1}
\end{equation}
Thus, the effective action for $L\to\infty$ becomes
\begin{eqnarray}
S_{\rm Ad} [u(\tau)] &&= \sum_{k=1}^{L} \big[
\frac{M}{2} 
\sum_j \frac{[ u_j(\tau_k) - u_j(\tau_{k-1}) ]^2}{\Delta \tau} 
\nonumber \\ 
&&+ \Delta \tau W(u(\tau_k)) \big] + i \theta [u(\tau)]
\;.\label{eq:s_eff}
\end{eqnarray}

Equivalent results can be derived in the Hamiltonian
approach to the adiabatic approximation. The basic idea
here is to restrict the full electron-lattice Hilbert
space to an ``adiabatic'' subspace which is spanned
by the set of 0th order adiabatic 
electron-lattice eigenstates
$\mid\chi_u^{(\kappa)}\rangle$ defined above in 
Eqs.~(\ref{eq:chi_u})-(\ref{eq:w_0})
with $\kappa$ restricted to the electronic
groundstate $\kappa=0$. The adiabatic subspace
thus consists of EP states of the general form
\begin{equation}
\mid \phi \rangle = \int d^Nu \;\; \phi(u) \mid \Psi_u^{(0)} \rangle
\label{eq:phi_ad}
\end{equation}
where $\phi(u)$ is an arbitrary (square-integrable) wavefunction 
amplitude which depends only on the lattice coordinates $u$.
The basic approximation step is then to
project the full EP Hamiltonian $H$ onto
the adiabatic subspace. In this manner one 
arrives at a 1st order effective
Hamiltonian $H_{\rm Ad}$ which is mathematically equivalent
to the 1st order adiabatic Euclidean action $S_{\rm Ad}$
in (\ref{eq:s_eff}), after $L\to\infty$. 
Since the adiabatic EP states $\mid \phi\rangle$
can be expressed entirely in terms of their
wavefunction amplitude $\phi(u)$, one can recast
$H_{\rm Ad}$ into the form of an effective Hamiltonian
acting only on the lattice coordinates $u$ in $\phi(u)$, 
without explicit reference to the underlying electronic
groundstate wavefunction $\mid\Psi_u^{(0)}\rangle$ 
contained in $\mid\phi\rangle$.
However, it is crucial to keep in mind the 
formal relationship (\ref{eq:phi_ad})
between the full adiabatic EP state $\mid \phi\rangle$
and its lattice wavefunction amplitude $\phi(u)$
if one wants to properly compare 1st order adiabatic results 
to exact results, obtained by e.g. numerically diagonalizing
the full EP Hamiltonian on small model 
clusters \cite{RaTh}.

In systems obeying standard harmonic lattice dynamics, the 
0th order Born-Oppenheimer ``energy surface'' $W_0(u)$ exhibits 
one unique global minimum configuration $u^{({\rm min})}$ which is, 
in terms of energy or in terms of configurational distance,
well separated from other, if existent, local minima. In that case, 
the path integral is dominated by small-amplitude ``harmonic''
fluctuations around $u^{({\rm min})}$ and a description of the 
lattice dynamics in terms of renormalized harmonic oscillators, 
i.e., phonons, remains valid. 
Since the displacement excursions around $u^{({\rm min})}$
are small, so are the fluctuations in the electronic 
wavefunction $\mid\Psi(u)\rangle$;
hence the small-amplitude (``phonon'') paths 
all have $\theta[u(\tau)]=0$ and Berry phase 
effects are negligible. Also, the $u$-derivatives
of $\mid\Psi(u)\rangle$ entering into $W_1$ are well behaved and 
the $m$-th order $u$-derivatives of the overlap
matrix elements 
$\langle \partial_{u_j} \Psi(u) \mid \partial_{u_j} \Psi(u) \rangle$
are typical of the order of inverse lattice constants
or inverse atomic distances raised to the $(m+2)$-th power.
The $W_1$-contribution to the harmonic restoring force constants,
for example, are thus smaller than the 0th order 
$W_0$-contributions by factors
of order of the 4th power of the lattice oscillator
zero-point displacement amplitude over the lattice constant.
Thus, the electronic overlap factor effects
$W_1(u)$ and $\theta[u(\tau)]$ can be altogether neglected.

By contrast, in polaronic systems
the 0th order lattice potential $W_0$ exhibits a large number
of nearly degenerate local minima. The low-energy
lattice dynamics is dominated by
tunneling processes between the local minima
which requires anharmonic large-amplitude
excursions of the local displacement coordinates $u_j$
and large local rearrangements of the electronic wavefunction
$\mid\Psi(u)\rangle$ \cite{Zhong}.
In that case electronic overlap
effects arising from both $\theta[u(\tau)]$ and
$W_1(u)$ can become quite important.

\section{TWO-SITE PROBLEM}
\label{sec:two-site}

The two-site version of the Holstein-Hubbard model \cite{EmHo,capone}
(\ref{eq:model}--\ref{eq:h_phon})
is a simple toy problem which retains some essential physical
features of the lattice polaron problem. We will use
it here to elucidate the basic underlying physical ideas
and formal concepts of
our adiabatic treatment of polaron formation 
and polaron tunneling
dynamics and, also, to
test the validity and illustrate some important limitations 
of the adiabatic approximation.
We restrict ourselves to the single-electron case 
on two sites, with an electronic intersite hybridization $t$.
Hence there are no correlation (Hubbard-$U$) effects
and the adiabatic electronic wavefunction $\mid\Psi(u)\rangle$
can be solved exactly by diagonalizing 
$H_0(u)$ which reduces to a $2\times 2$ matrix.

The two sites are labeled $1$ and $2$ with on-site
oscillator coordinates $u_1$ and $u_2$ and on-site
electron occupation numbers $n_1$ and $n_2$.
Introducing symmetrized displacement coordinates
\begin{equation}
u_\pm=(u_1 \pm u_2)/\sqrt{2}
\;,
\label{eq:u_pm}
\end{equation}
$W_0$ and $W_1$ can be written as
\begin{equation}
W_0(u) \equiv W_{0+}(u_+) + W_{0-}(u_-) \;,
\label{eq:w_0-2s}
\end{equation}
where
\begin{eqnarray}
W_{0+}(u_+) &=& \frac{K}{2}u_+^2 + \frac{C}{\sqrt{2}}u_+ 
\nonumber \\ &=&\left[ \frac{1}{2} \Big(\frac{u_+}{u_{\rm P}}\Big)^2 
+ \frac{1}{\sqrt{2}} \Big(\frac{u_+}{u_{\rm P}}\Big) \right]E_{\rm P}
\;,
\label{eq:w_0plus-2s}
\end{eqnarray}
\begin{eqnarray}
W_{0-}(u_-) &=& \frac{K}{2}u_-^2 - 
\sqrt{ \frac{C^2u_-^2}{2}+t^2 }
\nonumber \\ &=&\left[ \frac{1}{2} \Big(\frac{u_-}{u_{\rm P}}\Big)^2 
- \sqrt{ \frac{1}{2} \Big(\frac{u_-}{u_{\rm P}}\Big)^2 
          +\Big(\frac{t}{E_{\rm P}}\Big)^2 } \right]E_{\rm P}
\;,
\label{eq:w_0minus-2s}
\end{eqnarray}
and
\begin{equation}
W_1(u)\equiv W_{1-}(u_-)=
\frac{1}{4}\frac{\Omega^2}{E_{\rm P}} 
\frac
{
\Big( \frac{t}{ E_{\rm P} } \Big)^2
}
{
\left[ \Big(\frac{u_-}{u_{\rm P}}\Big)^2 
+2 \Big( \frac{t}{ E_{\rm P} } \Big)^2 \right]^2
}
\;.
\label{eq:w_1-2s}
\end{equation}
There is no Berry phase, i.e., $\theta[u(\tau)]\!\equiv\!0$, and the 
problem becomes equivalent to solving the Hamiltonian dynamics of 
a (fictitious) quantum particle of mass $M$ 
moving in a two-dimensional $(u_+,u_-)$-plane subject 
to the effective potential $W(u)=W_0(u)+W_1(u)$.
Because of (\ref{eq:w_0-2s}) and (\ref{eq:w_1-2s})
this dynamics is separable when written in terms of
$u_+$- and $u_-$-coordinates.

Since $u_+$ couples only to the total electron charge
$n_+\equiv n_1+n_2$, which is conserved, the $W_{0+}$-part of 
$W_0$ is just a harmonic potential, with its equilibrium position 
shifted to 
\begin{equation}
u_+^{(0)}= -u_{\rm P}/\sqrt{2}
\label{eq:u_+0}
\end{equation}
by the constant pulling force $Cn_+/\sqrt{2}$ exerted by the total
electron charge. Also, the electron groundstate
wavefunction $\mid\Psi(u)\rangle\equiv\mid\Psi(u_-)\rangle$, 
and hence $W_1$, does not depend on $u_+$.
The dynamics of the $u_+$-coordinate is therefore trivial,
at least for processes conserving the total electron number.

Since $u_-$ couples to the charge imbalance 
$n_-\equiv n_1-n_2$ between the two sites,
the shape of $W_{0-}$ is renormalized by the EP coupling
and $W_1$ contributes to the $u_-$-dynamics. We first consider
the 0th order contribution $W_{0-}$, 
shown in Fig.~\ref{fig:double_well}(a)
for several $E_{\rm P}$-values.

For small $E_{\rm P}$, $W_{0-}$ retains a single global minimum at 
$u_-=0$. Its harmonic restoring force
\begin{equation}
K_{0-} \equiv \frac{\partial^2}{\partial u_-^2} W_{0-}(u_-=0)
       = \Big( 1 - \frac{E_{\rm P}}{2t} \Big) K
\label{eq:k_0-}
\end{equation}
softens with increasing $E_{\rm P}$ and changes sign 
when $E_{\rm P}$ reaches a critical value
\begin{equation}
E_{\rm P}^{({\rm crit})} = 2 t
\;.
\label{eq:e_p-crit}
\end{equation}
For $E_{\rm P}>E_{\rm P}^{({\rm crit})}$, the character of $W_{0-}$ 
changes qualitatively: $W_{0-}$ acquires two degenerate minima
at $u_-=\pm u_{-}^{(0)}$, separated by a maximum at $u_-=0$, with
\begin{equation}
u_{-}^{(0)} = \sqrt{ 1 - \Big( \frac{2t}{E_{\rm P}}\Big)^2 } 
\frac{u_{\rm P}}{\sqrt{2}} 
\label{eq:u_-0}
\end{equation}
where $u_{\rm P}=C/K$ is the polaron shift (\ref{eq:u_pol}).
$u_{-}^{(0)}$ approaches $u_{\rm P}/\sqrt{2}$ 
in the strong-coupling limit
\begin{equation}
E_{\rm P}\gg t
\;.
\label{eq:strong_ep}
\end{equation}
The height of the 0th order potential 
barrier separating the two minima,
\begin{equation}
\Delta_{B0} \equiv W_{0-}(0)-W_{0-}(\pm u_-^{(0)}) =
\Big( \frac{1}{2} - \frac{t}{E_{\rm P}} \Big)^2 E_{\rm P}
\label{eq:delta_b0}
\end{equation}
increases with $E_{\rm P}$ and approaches $ E_{\rm P}/4 $
in the strong-coupling limit (\ref{eq:strong_ep}).

The physical origin of the double-well potential can be
most easily understood starting from the ``ionic'' ($t=0$)
limit of the model: 
For $t=0$, the two electronic eigenstates of $H_0(u)$,
\begin{equation}
\mid \Psi^{(\ell 1)} \rangle \equiv\mid n_1=1,\;\; n_2=0\rangle
\label{eq:psi-i}
\end{equation}
and 
\begin{equation}
\mid\Psi^{(\ell 2)}\rangle \equiv\mid n_1=0,\;\; n_2=1\rangle
\label{eq:psi-ii}
\end{equation}
have the electron completely localized on site $1$ and $2$, 
respectively, with eigenenergies
\begin{equation}
W^{(\ell 1,2)}(u_-)=\frac{K}{2}(u_- \pm u_{\rm P}/\sqrt{2})^2
- \frac14 E_{\rm P}
\;,
\label{eq:w-i-ii}
\end{equation}
where the upper (lower) sign refers to $W^{(\ell 1)}$ 
($W^{(\ell 2)}$), as shown by the two parabolic potential curves
in Fig.~\ref{fig:double_well}(a).
Assuming $C>0$, $\mid\Psi^{(\ell 1)}\rangle$ is the groundstate 
for $u_-<0$ and $\mid\Psi^{(\ell 2)}\rangle$ for $u_->0$. At $u_-=0$,
the two parabolic eigenenergy curves $W^{(\ell 1)}(u_-)$ 
and $W^{(\ell 2)}(u_-)$ intersect, both states
are degenerate and the groundstate wavefunction changes 
discontinuously as a function of $u_-$. 
When the hybridization $t$ is turned on, 
the two fully localized wavefunctions $\mid\Psi^{(\ell 1)}\rangle$
and $\mid\Psi^{(\ell 2)}\rangle$ become mixed, 
the electronic degeneracy at $u_-=0$ is lifted 
and a minimum excitation gap of $2t$ opens up between the
electronic ground- and 1st excited states. The sharp cusp 
at $u_-=0$ in the $t=0$ double-parabolic potential function
\begin{eqnarray}
W_{0-}(u_-)\mid_{t=0}&&=\min(W^{(\ell 1)}(u_-), W^{(\ell 2)}(u_-)) 
\nonumber\\
&&= \frac{K}{2}(\mid u_-\mid-u_{\rm P}/\sqrt{2})^2
- \frac14 E_{\rm P}
\label{eq:w_0t0}
\end{eqnarray}
[see Fig.~\ref{fig:double_well}(a)] is 
rounded by the finite $t$; as a function of
$u_-$, the groundstate wavefunction $\mid\Psi(u_-)\rangle$
now changes continuously at $u_-=0$.
However, $\mid\Psi(u_-)\rangle$ still has predominantly
$\mid\Psi^{(\ell 1)}\rangle$-character near 
$u_-=-u_{-}^{(0)}$ and predominantly
$\mid\Psi^{(\ell 2)}\rangle$-character near $u_-=u_{-}^{(0)}$.
With increasing $t$, the tunneling barrier height decreases,
initially by about $t$. The barrier vanishes when $t$
reaches $t^{({\rm crit})}=E_{\rm P}/2$
which is equivalent to the above condition
(\ref{eq:e_p-crit}), for $E_{\rm P}^{({\rm crit})}$.

If one examines the groundstate wavefunction $\mid\Psi(u_-)\rangle$ 
and its charge distribution 
$\langle\Psi(u_-)\mid n_j\mid\Psi(u_-)\rangle$
for $E_{\rm P}>E_{\rm P}^{({\rm crit})}$, 
near the two potential minima 
$\pm u_{-}^{(0)}$, one thus finds the electron predominantly 
localized at site $1$ when $u_-\cong-u_{-}^{(0)}$
and predominantly at site $2$ when $u_-\cong +u_{-}^{(0)}$, assuming 
again $C>0$ here and in the following.
By contrast, at the potential minimum $u_-=0$
in the regime $E_{\rm P}<E_{\rm P}^{({\rm crit})}$, 
the electron charge is delocalized evenly
between the two sites $1$ and $2$. 
Thus, at the level of the 0th order adiabatic approximation
(i.e., neglecting the lattice kinetic energy),
the transition from the single-well potential
case $E_{\rm P}<E_{\rm P}^{({\rm crit})}$ to the double-well case 
$E_{\rm P}>E_{\rm P}^{({\rm crit})}$
is essentially a transition from a delocalized non-degenerate
groundstate ($u_-=0$)
to a localized degenerate groundstate ($u_-=\pm u_{-}^{(0)}$). 
In the former case, the electron's delocalization energy
dominates and it is energetically favorable for the electron
wavefunction to be spread out between the two sites; in the latter
case, the EP coupling dominates and favors localizing the electron 
charge on only one of the two sites. The lattice spontaneously 
distorts so as to set up an attractive EP ``potential well''
which binds and localizes the electron. The electronic binding
energy thus gained in turn stabilizes the local lattice distortion.
This self-localization mechanism is the essence of polaron formation.

Localizing the electron on either one
of the two sites is energetically equivalent due to the
reflection symmetry ($(1,2)\to(2,1)$) of the underlying
Hamiltonian. At the level of the 0th order adiabatic approximation,
this symmetry is broken in the 2-fold degenerate 0th 
order groundstates $u_-=\pm u_{-}^{(0)}$.
The existence of two {\it degenerate} local minima
in $W_{0-}$ can thus be understood as a direct consequence of the 
symmetry breaking which accompanies the self-localization transition.
In the 1st order adiabatic approximation, the lattice kinetic energy 
restores this symmetry by inducing tunneling processes between the
two potential minima, thus giving rise to
a non-degenerate groundstate in which the two degenerate
0th order states are admixed with equal probability 
weight.
% \cite{foot_finite_T}.

From the above discussion it is clear that such
tunneling processes are accompanied by a
transfer of electron charge between the two sites. 
These {\it lattice} tunneling processes, occurring within
the EP-induced multiple-well Born-Oppenheimer potential, 
constitute the basic low-energy mechanism whereby 
self-localized electrons can move through the lattice.
At higher temperatures, thermally activated
over-the-barrier hopping will dominate the 
polaron transport 
\cite{Holstein,EmHo}
which, again, can be described
as a purely lattice dynamical phenomenon.
Thus, within the framework of the 1st order 
adiabatic approximation, polaron
formation and polaron dynamics is fundamentally
reduced to a problem of {\it non-linear lattice}
dynamics. 

We now turn to the 1st order potential correction $W_{1-}(u_-)$
(\ref{eq:w_1-2s}) in the two-site problem, shown for several
values of $t/E_{\rm P}$ in Fig.~\ref{fig:double_well}(b). 
Since $W_1(u)$, according to (\ref{eq:w_1}), 
is controlled by the $u$-gradient
of the electron wavefunction 
$\mid\Psi(u)\rangle$, we should expect it to
exhibit peaks wherever $\mid\Psi(u)\rangle$ varies most
rapidly with $u$. In the two-site problem, this occurs at
$u_-=0$ where $\mid\Psi(u)\rangle$ changes its character from being
predominantly localized on site $1$ to being localized
on site $2$, as discussed above. For large $\mid u_-\mid$,
$\mid\Psi(u_-)\rangle$ approaches a constant, either 
$\mid\Psi^{(\ell 1)}\rangle$
or $\mid\Psi^{(\ell 2)}\rangle$, 
hence $W_{1-}(u_-)\to 0$ for $\mid u_-\mid\to\infty$.
%According to ({\ref{eq:w_1-2s}), the peak height at $u_-=0$,
%%
%\begin{equation}
%\Delta_{B1} \equiv W_1(0) = ZZZ\Big\frac{\Omega}{t}\Big)^2 E_{\rm P}
%\;,
%\label{eq:delta_b1}
%\end{equation}
%%
%varies as $(\Omega/t)^2E_{\rm P}$
%and the peak width (FWHM)
%%
%\begin{equation}
%\Delta u_1 = ZZZ  \frac{t}{E_{\rm P}} u_{\rm P}
%\label{eq:delta-u_1}
%\end{equation}
%%
%is linear in $t/E_{\rm P}$.
%
In the polaron regime $E_{\rm P}>E_{\rm P}^{({\rm crit})}$, 
the primary effect of $W_1$ 
is to enhance the tunneling barrier separating the two
potential minima. In addition, $W_{1-}(u_-)$ will also tend to shift 
the two polaronic potential minima further apart,
%in the total $u_-$-potential $W_-\equiv W_{0-}+W_1$,
thus causing the tunneling barrier to become wider than
in the 0th order potential $W_0$. Both of these $W_1$-effects
tend to suppress the tunneling rate through the barrier.
Even though $W_1(0)$ may be small compared
to the 0th order barrier height $\Delta_{B0}$ (\ref{eq:delta_b0}),
its effect on the polaron tunneling rates can be
quantitatively of some importance,
since tunneling rates are typically
exponentially sensitive to changes 
in the tunneling barrier.

In the delocalized regime $E_{\rm P}<E_{\rm P}^{({\rm crit})}$,
the primary effect of $W_1$ is to soften the harmonic restoring
force constant $K_-\equiv \partial_{u_-}^2 W_-(0)$ by an amount
\begin{eqnarray}
K_{1-} &&\equiv \frac{\partial^2}{\partial u_-^2} W_{1-}(u_-=0)
\nonumber \\
       &&= - \frac18 \Big(\frac{\Omega}{t}\Big)^2 
               \Big(\frac{E_{\rm P}}{t}\Big)^2 K < 0
\;.
\label{eq:k_1-}
\end{eqnarray}
Thus, $W_1$ also lowers the critical $E_{\rm P}$ for the on-set
of polaron formation. However, in the large-$M$ limit 
where the adiabatic approximation is valid, that is, 
for $\Omega\ll t$ (see below), these corrections are smaller
than the 0th order results (\ref{eq:k_0-},~\ref{eq:e_p-crit}) 
by factors of order $(\Omega/t)^2\times(E_{\rm P}/t)^2$.
Provided $t$ and $E_{\rm P}$ are of comparable magnitude
and $\Omega \lesssim t$ (see below),
$W_1$ does not qualitatively
alter the basic structure of the lattice potential $W$
in either coupling-strength regime.
However, $W_1$ can become qualitatively important
in suppressing certain non-adiabatic 
processes in lattice systems,
as will be discussed in the next section.

\section{VALIDITY AND LIMITATIONS OF THE ADIABATIC APPROXIMATION}
\label{sec:validity}

The basic criterion for the validity of the adiabatic
approximation is that the longest time scale of
the electronic motion should be short compared to the
shortest time scale of the lattice motion or, equivalently,
the lowest electronic frequency scale should be large
compared to the highest lattice frequency scale.
In the two-site problem, the lowest electronic
frequency scale is the excitation energy between the
electronic groundstate $\mid\Psi(u)\rangle$ and the 1st excited
state which is at least $2t$ (at $u_-=0$) or larger.
The highest lattice frequency scale is the phonon
energy $\Omega$ and hence we expect the adiabatic
approximation to work, provided that
\begin{equation}
\Omega \ll 2t
\;.
\label{eq:valid}
\end{equation}

In the polaron regime $E_{\rm P}>E_{\rm P}^{({\rm crit})}$, 
the lattice (not the electron !) motion acquires an additional, 
low frequency scale, given by the polaronic tunneling splitting 
$2t_P$ between the ground- and 1st excited states in 
the double-well lattice potential $W(u)$. 
This tunneling energy scale
is typically smaller than or, at most, comparable to
the bare phonon energy scale $\Omega$, 
given the conditions where a polaronic
double-well forms in the first place. 
Hence, the basic criterion (\ref{eq:valid})
applies in the polaronic regime just as well
as in the delocalized regime, regardless of the
electron-phonon coupling strength. Criterion (\ref{eq:valid})
applies even in the strong-coupling regime 
(\ref{eq:strong_ep}) where $2t_P$ becomes orders of 
magnitude smaller than $\Omega$.

Although the foregoing point has been established
for some 30 years now \cite{Holstein,EmHo,capone}, we wish
to strongly re-emphasize it here because
a great deal of confusion about this 
has been created in the more recent literature
on the two-site problem, as for example in 
Ref.~\onlinecite{RaTh}.
The basic error in some of the more recent 
work is to regard the polaron tunneling splitting 
$2t_P$, rather than $2t$, as the lowest
relevant electronic energy scale. Doing so, one then arrives
at the much too restrictive validity criterion
\begin{equation}
\Omega \ll 2t_P
\;.
\label{eq:wrong}
\end{equation}
If correct, this would imply that the polaron regime
$E_{\rm P}>E_{\rm P}^{({\rm crit})}$ can not be treated
in the adiabatic approximation, since typically 
$t_P\lesssim \Omega$ even under the most favorable
conditions. In the strong-coupling regime
(\ref{eq:strong_ep}) where $t_P \ll \Omega$
the adiabatic approximation should break down
completely according to (\ref{eq:wrong}).

The fundamental misconception here is that $2t_P$
is of course {\it not} the lowest {\it electronic}
energy scale, but rather represents an energy scale
of the {\it lattice} motion, 
%moving subject to the strongly renormalized
%double-well effective potential $W$, 
as discussed above. The relevant lowest electronic 
energy splitting, between the electronic
ground- and 1st excited states
{\it at fixed lattice coordinate} $u$
is at least $2t$ in the two-site model,
regardless of whether $E_{\rm P}$ is small or large.

To illustrate this point, we have generated exact 
numerical solutions of the two-site problem
using the full Hamiltonian $H$ without any approximation, 
and compared to solutions of the 1st order
effective adiabatic effective Hamiltonian 
$H_{\rm Ad}\equiv H_{\rm M}+W$, corresponding
to the effective action $S_{\rm Ad}$ from (\ref{eq:s_eff}).
For both the exact and the adiabatic problems, we have used
a sufficiently fine discretization of the $u_-$ coordinate
and a sufficiently large cut-off at large $u_-$ to ensure
a numerical accuracy of better than $1\%$ in the calculated
energy splittings over the entire parameter range
studied. In Fig.~\ref{fig:tunnel2site}, we show the logarithm
of the polaron tunneling splitting $2t_P$, 
that is, the excitation energy from the ground- to the 
1st excited states of the full electron-phonon system, 
as a function of $E_{\rm P}/\Omega$ for $t\equiv 1$ and four 
different EP couplings, $E_{\rm P}=2.5$, $3$, $4$ and $8$ 
which are well inside the polaronic regime 
($E_{\rm P}>E_{\rm P}^{({\rm crit})}$).

In addition to the exact solution,
we show two different adiabatic solutions in 
Fig.~\ref{fig:tunnel2site}, one obtained
with the full adiabatic lattice potential $W\equiv W_0+W_1$,
the other using only the 0th order potential, $W\cong W_0$.
These are being referred to in the following
as the ``full'' and as the ``simple''
adiabatic solutions, respectively. As expected from 
the Holstein-Lang-Firsov strong-coupling 
expansion \cite{Holstein,capone,LaFi}
and from semi-classical (WKB) arguments,
the tunneling splitting at fixed $E_{\rm P}$ and 
$t$ decreases exponentially with $1/\Omega$, 
as indicated by a roughly linear dependence of
$\ln (2t_P)$ on $1/\Omega$ in Fig.~\ref{fig:tunnel2site}.

Remarkably, the full adiabatic result agrees with the
exact solution to better than $14\%$ over a parameter region
$0.15t<\Omega<0.5t$ wherein $2t_P$ varies by more than 9 
orders of magnitude, including the regime where $2t_P$
is orders of magnitude smaller than $\Omega$. The simple
($W\cong W_0$) adiabatic solution reproduces the qualitative features
of the $1/\Omega$- and $E_{\rm P}$-dependence of $2t_P$
quite well, but the quantitative agreement is noticeably
worse than for the full adiabatic solution.
The agreement between the full adiabatic 
and the exact results is all the
more convincing in light of the fact that the tunneling
splitting is ``exponentially sensitive'' to small 
errors or changes in the wavefunction inside the tunneling barrier.
Thus, our comparison of the tunneling splittings 
constitutes a much more stringent test of the underlying
approximations than a comparison of, say,
low-lying-state expectation values or wavefunction
amplitudes. Other exact numerical results for the two-site problem,
such as reported e.g. in Ref.~\onlinecite{RaTh}, are generally
in equally good agreement with the corresponding adiabatic
solution, provided, that is, one exercises enough care 
to use the proper adiabatic wavefunctions
$\mid\phi\rangle$, Eq.~(\ref{eq:phi_ad}), in carrying out 
the comparison.

As expected from (\ref{eq:valid}), the agreement between adiabatic
and exact results deteriorates at high phonon frequencies
when $\Omega$ becomes comparable to $t$. As a practical 
matter, even for $\Omega\cong 2t$, the agreement is still quite
acceptable. For applications to lattice systems,
it is of interest to explore in some detail how 
the adiabatic approximation actually breaks down as one enters
into the ``anti-adiabatic'' regime
\begin{equation}
t\ll\Omega
\;.
\label{eq:anti_ad}
\end{equation}
As a limiting case, we consider the ionic 
limit $t\to 0$, already discussed above. Here, the
Holstein-Hubbard problem can be trivially solved exactly
\cite{SZF}. Obviously there cannot
be any electron tunneling between the two
sites and the polaron tunneling splitting $2t_P$ 
vanishes.

By contrast, in the simple adiabatic approximation
$W\cong W_0$, $W_{0-}$ approaches the double-parabolic
potential (\ref{eq:w_0t0})
for $t\to0$, which has a tunneling 
barrier of finite height and width.
The simple adiabatic approximation would thus predict
a non-vanishing finite tunneling splitting $2t_P>0$
even for $t=0$, a clearly unphysical result.

If instead one uses the full adiabatic approximation,
with $W=W_0+W_1$, the correct qualitative
physical behavior of $2t_P$ is restored by the $W_1$-term,
shown in Fig.~\ref{fig:double_well}(b):
According to Eq.(\ref{eq:w_1-2s})
%Eqs.(\ref{eq:delta_b1}-\ref{delta-u_1}),
the $W_1$-peak height (at fixed $E_{\rm P}$ and $\Omega$)
diverges as $t^{-2}$, while at the same time 
its peak width vanishes, but only 
linearly in $t$ in the limit $t\to0$. It is then easy to show
that the transmission amplitude through the 
$W_1$-barrier vanishes, that is, the barrier becomes 
impenetrable in the limit $t\to0$ which forces 
$2t_P\to0$ for $t\to0$.
%The 1st order adiabatic predictions for $2t_P$ in the
%small-$t$ limit (\ref{eq:anti_ad}) differ from the exact
%results quantitatively, by factors of $2-3$ or more,
%but they generally do exhibit the correct qualitative
%$E_{\rm P}$- and $\Omega$-dependence, predicted e.g.
%by the Holstein \cite{Holstein,LaFi}
%strong-coupling expansion. 
Thus, as far as the tunneling
splitting $2t_P$ is concerned, the full adiabatic approximation 
reproduces qualitatively the correct physical behavior
even in the extreme anti-adiabatic regime.

The actual failure of the full adiabatic approximation in the 
$t\to 0$-limit is a more subtle problem. 
It consists of the unphysical constraint being imposed 
on the dynamics of the $u_-$-coordinate by the
impenetrability of the $W_1$-barrier. For $t\to0$, the 
$W_1$-barrier forces the lattice wavefunction $\phi(u)$ in
(\ref{eq:phi_ad})
to vanish identically either to the right ($u_->0$) 
or to the left ($u_-<0$) of the barrier. Thus,
the amplitude for propagation from an initial $u_-<0$
to a final $u_->0$ (or reverse) vanishes in 
the full adiabatic approximation at $t=0$.
In the exact solution of the $t=0$ problem,
this constraint does not exist; the lattice is free to propagate
with some finite amplitude from $u_-<0$ to $u_->0$.
In the exact $t=0$ solution, the lattice dynamics is
governed either by the left or 
the right parabolic well, 
$W^{(\ell 1)}$ or $W^{(\ell 2)}$, corresponding respectively to the 
left-localized or to the right-localized electron states,
$\mid \Psi^{(\ell 1)}\rangle$ or $\mid \Psi^{(\ell 2)}\rangle$, 
discussed in Sec.~\ref{sec:two-site}. 
The problem with the adiabatic approximation is that
the $t=0$ electron groundstate $\mid \Psi(u)\rangle$ 
exhibits a level crossing and thus changes discontinuously
at $u_-=0$, as discussed above. The adiabatic approximation,
by construction, excludes transitions between, say, the
electronic ground- and 1st excited eigenstates.
But this is just what happens at $u_-=0$ in the $t\to0$-limit:
If the lattice coordinate crosses $u_-=0$ from the left, say,
under exact time evolution, the electron remains in its localized 
state $\mid\Psi^{(\ell 1)}\rangle$, which is the groundstate only 
for $u_-<0$, but becomes the 1st excited state when $u_->0$.
The adiabatic approximation on the other hand
forces the electron to remain in the groundstate
which changes discontinuously at $u_-=0$, from 
$\mid\Psi^{(\ell 1)}\rangle$ to $\mid\Psi^{(\ell 2)}\rangle$.

In the two-site problem, the foregoing impenetrability
constraint causes only a small error, of order 
$\exp(-E_{\rm P}/\Omega)$,
in the low-lying lattice eigenstates and energies 
if the lattice oscillator zero-point amplitude 
is small compared
to the double-well separation $\sqrt{2}u_{\rm P}$, 
that is, if $\Omega\ll E_{\rm P}$.
However, the impenetrability constraint 
may introduce a qualitative failure
of the adiabatic approximation
if applied to large systems $N\to\infty$
and tunneling processes which 
transfer a polaron in a single step
over large distances, as we will now discuss.

Let us consider for simplicity the 
case of the Holstein model for
just a single electron in a large lattice 
with sufficiently strong $E_{\rm P}$ to form a polaron. 
Suppose the polaron is
localized at some site $\xi$, say, and we want to study the
tunneling barrier for transferring the polaron 
in a single tunneling step to a distant
site $\zeta=\xi+r$, i.e., with $|r|\gg a$ where $a$
is the lattice constant.

Let $u^{(\xi)}\equiv(u^{(\xi)}_1\dots u^{(\xi)}_N)$ 
denote that lattice configuration which 
minimizes $W_0(u)$ and localizes
the polaron around the ``centroid site'' $\xi\in\{1\dots N\}$. 
That is, $|u^{(\xi)}_\ell|$ and the
corresponding electron charge density 
$\langle n_\ell\rangle^{(\xi)}$ are 
maximal at $\ell=\xi$ and die out exponentially 
at large distances $|\ell-\xi|$ from the centroid.
Likewise, let $u^{(\zeta)}$ denote the lattice configuration
which localizes the polaron around site $\zeta$. By 
lattice translational invariance
\begin{equation}
u^{(\zeta)}_\ell= u^{(\xi)}_{\ell-r}
\;.
\label{eq:latt_trans}
\end{equation}
if $\zeta=\xi+r$. Notice that polaron formation
breaks the translational symmetry of the lattice
in the $0th$ order adiabatic approximation.
As a consequence, $W_0$ exhibits $N$ degenerate
local minima, corresponding to the $N$ different,
but translationally equivalent
$u^{(\xi)}$ configurations on an $N$-site lattice
with periodic boundary conditions.
This is the lattice analogue to the breaking of 
reflection symmetry in the two-site problem.

Let $u^{(\zeta\xi)}(s)$ denote the linear path segment
in the $N$-dimensional $u$-space connecting 
$u^{(\xi)}$ to $u^{(\zeta)}$, i.e.,
\begin{equation}
u^{(\zeta\xi)}(s) = \Big({1\over2}-s\Big) u^{(\xi)} 
             +\Big({1\over2}+s\Big) u^{(\zeta)}
\label{eq:tun_path}
\end{equation}
with $s\in [-{1\over2},+{1\over2}]$. 
In the following discussion, we consider 
(\ref{eq:tun_path}) as a representative of
low-action tunneling trajectories
connecting $u^{(\xi)}$ to $u^{(\zeta)}$.
The $s$-coordinate can thus be regarded as
the lattice analogue to the $u_-$ tunneling coordinate 
(\ref{eq:u_pm}) in the two-site problem.
Note in particular that $W_0(u^{(\zeta\xi)}(s))$
has local minima at $s=-{1\over2}$ and $s=+{1\over2}$
which must, by continuity, be separated
by (at least) one intervening maximum, i.e., by
a tunneling barrier. The simplest scenario, normally
borne out in the numerical calculations
discussed below, is that there is only one
barrier maximum, by symmetry located at $s=0$.
Thus, along $u^{(\zeta\xi)}(s)$, $W_0$ has qualitatively 
the same structure as $W_{0-}(u_-)$ described above
for the two-site problem.

The first crucial point to note here
is that the width of this tunneling barrier,
that is, the Euclidean distance from $u^{(\xi)}$ to $u^{(\zeta)}$
in their $N$-dimensional $u$-space, 
\begin{eqnarray}
d(\zeta,\xi) &&\equiv |u^{(\zeta)}-u^{(\xi)}| 
\nonumber\\
&&\le |u^{(\zeta)}|+|u^{(\xi)}| = 
2|u^{(\xi)}|\equiv d_\infty
\label{eq:udist}
\end{eqnarray}
is finite and bounded by an upper limit
$d_\infty$ which is independent of the
spatial distance $|\zeta-\xi|=|r|$. Note that
$d_\infty$ is independent of $\xi$ or $\zeta$
due to lattice translational invariance.
Thus two polaron configurations $u^{(\xi)}$ and
$u^{(\zeta)}$ are never further apart from each other
than $d_{\infty}$ in $u$-space, regardless of
how far apart their centroid sites $\xi$ and $\zeta$ 
are in real space.

The second important point is that 
the height of the 0th order ($W_0$)
tunneling barrier along $u^{(\zeta\xi)}(s)$
is also bounded independently of lattice 
distances $|\zeta-\xi|$.
To see this, note that the EP potential
$C u_\ell$ acting on the electron
is attractive, i.e., $Cu_\ell<0$,
%tends to lower the electron
%groundstate energy, compared to the undistorted
%lattice configuration) 
for any $u$-configuration along the path 
$u^{(\zeta\xi)}(s)$ between $s=0$ and $s=1$.
Hence, the contribution to $W_0(u)$ from $H_{\rm e}+H_{\rm e-ph}(u)$
is bounded from above by the electron
groundstate energy of the undistorted lattice.
Also, by an argument analogous to (\ref{eq:udist}),
the elastic energy contribution $H_{\rm K}(u)$
is bounded from above by ${3\over2}H_{\rm K}(u^{(\xi)})$.
Both of these upper bounds are independent of $|\zeta-\xi|$.
%This completes the proof that $W_0(u)$ is bounded
%along $u^{(\zeta\xi)}(s)$ by an upper limit which is independent
%of the spatial distance $|\zeta-\xi|$ between the two 
%centroid sites. 

The foregoing considerations suggest that a
manifold of tunneling trajectories exists, sufficiently close to
$u^{(\zeta\xi)}(s)$, which will all connect $u^{(\xi)}$ to
$u^{(\zeta)}$ through a $W_0$-barrier whose height
and width is bounded by upper limits 
independent of $|\zeta-\xi|$.
Within the simple adiabatic approximation, $W=W_0$,
one thus arrives at the unphysical result
that the polaron can tunnel in a single
(``instanton'') tunneling step from any site $\xi$ to any
site $\zeta$ in the lattice with a tunneling
matrix element $t_P(\zeta-\xi)$ which does {\it not} 
go to zero for $|\zeta-\xi|\to\infty$, but rather
\begin{equation}
\lim_{|\zeta-\xi|\to\infty} |t_P(\zeta-\xi)| \equiv t_{P\infty} > 0
\;.
\label{eq:t_p-infty}
\end{equation}
The foregoing argument can be made formally more rigorous,
by employing instanton methods similar to those
described in the next section for short-distance 
tunneling processes.
We will not engage in that exercise here. Suffice it to
say that the simple adiabatic result (\ref{eq:t_p-infty})
for the lattice is analogous to the above described
two-site result in the $t=0$ limit: 
the simple adiabatic approximation allows tunneling
solely on the basis of the $W_0$
electronic groundstate {\it energy} barrier,
regardless of whether there is actually any 
electronic {\it wavefunction overlap} between the initial
and final $u$-configurations of the tunneling process.

To account for wavefunction overlap effects in long-distance
tunneling processes, the $W_1$-term (\ref{eq:w_1}) 
has to be included in the total potential $W=W_0+W_1$.
Let us consider the evolution of the
electronic groundstate wavefunction  
$\mid \Psi(u) \rangle$
along the linear tunneling trajectory $u^{(\zeta\xi)}(s)$
(\ref{eq:tun_path})
between two centroid sites $\zeta$ and $\xi$
with $|\zeta-\xi|\gg \ell_P(u)$. Here $\ell_P(u)$ denotes
the exponential localization length of $\mid \Psi(u)\rangle$
for local lattice distortions comparable to $u^{(\xi)}$.
As a simplest scenario, let us assume
that the wavefunction $\mid\Psi(u)\rangle$ remains
localized for all $u$ along $u^{(\zeta\xi)}(s)$.
This situation will be realized at
EP coupling strengths $E_{\rm P}$ which are sufficiently
large compared to $E_{\rm P}^{({\rm crit})}$. The electronic
groundstate problem can then be qualitatively described
as follows:

The EP potential $Cu^{(\zeta\xi)}_\ell(s)$, 
acting on the electron at sites $\ell$,
consists of two localized wells, 
$C({1\over2}-s)u^{(\xi)}_\ell$ 
and 
$C({1\over2}+s)u^{(\zeta)}_\ell$, the former centered around 
site $\ell=\xi$,
the latter around $\ell=\zeta$. 
As $s$ is varied from $-{1\over2}$ to $+{1\over2}$,
the EP well at $\xi$ becomes shallower and the EP well
at $\zeta$ deepens. At $s=0$, the two wells become degenerate.
Assuming large real-space tunneling distances 
$|\zeta-\xi|$, the electron
wavefunction overlap between these two wells
is exponentially small.
Hence, the electron groundstate wavefunction
$\mid\Psi(u^{(\zeta\xi)}(s))\rangle$ will remain 
localized at site $\xi$ for most $s<0$ until $s$ gets 
very close to $s=0$.
Within a very small interval around $s=0$,
$\mid\Psi(u^{(\zeta\xi)}(s))\rangle$ 
will then switch over from being localized
around $\xi$ to being localized around $\zeta$. In that narrow
$s$-region around $s=0$,
the electron wavefunction consists of the superposition of two almost
non-overlapping localized parts, one centered around $\xi$, the
other around $\zeta$. Since $\mid \Psi(u)\rangle$ changes very rapidly
as a function of $u$ near $u^{(\zeta\xi)}(0)$, $W_1(u)$ 
will exhibit a sharp peak along $u^{(\zeta\xi)}(s)$
which increases the tunneling barrier at $s=0$ and hence
suppresses the tunneling amplitude.

Formally, this problem can be treated by a tight-binding ansatz
for the electron groundstate wavefunction:
$\mid\Psi(u^{(\zeta\xi)}(s))\rangle$ 
near $s=0$ is approximated by a superposition of
$\mid \Psi(u^{(\xi)}/2) \rangle$ and 
$\mid \Psi(u^{(\zeta)}/2) \rangle$,
i.e., by the single-well groundstates of the two EP wells
${1\over2}Cu^{(\xi)}_\ell$ and
${1\over2}Cu^{(\zeta)}_\ell$, discussed above.
As $s$ is varied near $s=0$,
the response of $\mid \Psi(u^{(\zeta\xi)}(s))\rangle$ to 
the changing EP well depths is then governed
by the effective {\it electronic} hybridization overlap
\begin{eqnarray}
t_{\rm eff}(\zeta-\xi) &&=
\langle \Psi(u^{(\xi)}/2)\mid H_{\rm e} 
\mid \Psi(u^{(\zeta)}/2) \rangle
\nonumber \\
&&\sim t \exp \Big( -2 { |\zeta-\xi| \over \ell_{P,{1\over2}} }\Big)
\label{eq:t_eff-mj}
\end{eqnarray}
where $\ell_{P,{1\over 2}}\equiv \ell_P(u^{(\xi)}/2)$
is the localization length of
$\mid \Psi(u^{(\xi)}/2) \rangle$.
Within the tight-binding ansatz, the problem
then becomes analogous to the two-site problem
in the $t\to 0$ limit, with the tight binding-basis states 
$\mid \Psi(u^{(\xi)}/2) \rangle$ 
and $\mid \Psi(u^{(\zeta)}/2) \rangle$
replacing the two-site basis states
$\mid\Psi^{(\ell 1)}\rangle$ and $\mid\Psi^{(\ell 2)}\rangle$,
respectively. $W_1(u^{(\zeta\xi)}(s))$ exhibits a sharply peaked 
barrier at $s=0$, analogous to the $t\to0$ limit of 
the two-site problem.
The $W_1$-barrier will be 
roughly of the form given by Eq.~(\ref{eq:w_1-2s}),
with $u_-$ replaced by $u_-(s)\equiv d(\zeta,\xi)\times s $ 
and with $t$ replaced by $t_{\rm eff}(\zeta-\xi)$. 
Thus, along with $t_{\rm eff}(\zeta-\xi)$,
the transmission amplitude through the $W_1$-barrier
and the effective polaron tunneling matrix element $t_P(\zeta-\xi)$
will decrease exponentially with the tunneling
distance $|\zeta-\xi|$, analogous to
the $t\to 0$-limit in the two-site problem.

The long-distance
polaron tunneling processes are in the anti-adiabatic
regime, since the relevant effective electronic
hybridization overlap matrix elements $t_{\rm eff}(\zeta-\xi)$
become small compared to the phonon energy $\Omega$
at large tunneling distances $|\zeta-\xi|$ on
large lattice sizes $N$. The $W_1$ potential
ensures, at least qualitatively, 
that the effective polaron tunneling matrix
elements $t_P(\zeta-\xi)$ are properly suppressed to zero
at large tunneling distances. Hence, the full
adiabatic approximation $W=W_0+W_1$ restores
the correct long-distance behavior, as far as the
polaron tunneling matrix element is concerned.

However, just as in the anti-adiabatic limit of the
two-site problem, the $W_1$-term also imposes
an unphysical constraint on the lattice coordinates.
In the present case, involving long-distance
tunneling on a lattice, this constraint acts to couple the
lattice displacement coordinates at arbitrarily
large distances $|\zeta-\xi|$, thereby introducing
unphysical infinite-range interactions between the
lattice coordinates.
%This could very well lead to potentially
%serious errors if one were trying to use the full
%adiabatic approximation to study the long-wavelength lattice dynamics.
%Further, mathematically more rigorous investigations 
%of this problem, especially in the correlated many-electron
%case, are warranted.

Thus, in long-distance tunneling processes, the preconditions
for the adiabatic approximation break down. However,
from the foregoing discussion it is clear that the
effective action for the corresponding paths increases exponentially
and that the corresponding tunneling matrix element dies out 
exponentially with the tunneling distance. The simplest
way of dealing with such long-distance
tunneling processes is therefore to altogether neglect the
corresponding tunneling paths in the path integral.
This is what we will do in the following analysis.
As far as the polaron tunneling dynamics is concerned,
the short-distance processes will be dominant.
The relevant effective electronic matrix elements $t_{\rm eff}$ 
for short-distance processes are of order of the 1st neighbor $t$ 
which is normally larger than or at least comparable to the phonon 
energy scale in typical solid state situations.
We {\it can} therefore use the adiabatic approximation 
to accurately estimate the effective action for short-distance
tunneling paths. And it is only in this limited sense that
the adiabatic approximation {\it will} be used in the following.

\section{INSTANTONS AND EFFECTIVE HAMILTONIAN}
\label{sec:instanton}

The problem of polaron formation in the 2D Holstein-$ tJ $
and Holstein-Hubbard models has already been studied
extensively.\cite{YBL,Zhong,Roeder,Holstein}
In the nearly ${1\over2}$-filled band regime, the 
dopant induced hole carriers in the AF spin background
can form polarons with much less EP coupling
strength than is required for a single electron in an empty band.
Thus $E_{\rm P}^{({\rm crit})}$ for forming a single hole polaron 
in the ${1\over2}$-filled system is reduced by a factor of about
$4-5$, compared to a single electron polaron formation
in the empty band system. This reduction in $E_{\rm P}^{({\rm crit})}$
has been explained in terms of the hole mass enhancement
and self-localization effect
in the AF spin background of the nearly ${1\over2}$-filled
Hubbard system \cite{Zhong}.
The basic idea here is that the coupling to the AF spin background
already provides some form of self-localization of the
hole relative to a self-induced local distortion of 
the AF spin correlations \cite{Zhong,ChSc}.
This spin polaron effect is manifested in the strongly
reduced hole quasiparticle bandwidth, from  $8t$ in the 
non-interacting system to $\sim 2J$ in Hubbard or $ tJ $ systems
near half-filling. In the presence of EP
coupling, this electronic bandwidth reduction
permits the hole quasi-particle to become self-trapped by
a much weaker EP potential well; hence the reduction
in $E_{\rm P}^{({\rm crit})}$. 
The fact that the polaron formation threshold
$E_{\rm P}^{({\rm crit})}>0$ remains non-zero
even in the strongly correlated systems
is dictated by the so-called 
small-polaron dichotomy \cite{emin_holstein_scaling},
as discussed further in Sec.~\ref{sec:p_liquid}.

For a multi-hole system containing 
\begin{equation}
P \equiv N - \sum_j n_j
\label{eq:p_hole}
\end{equation}
doping-induced holes on an $N$-site lattice, there are 
(\raise0.9ex\hbox{$N$}\kern-0.9em\lower1.0ex\hbox{$P$})
possible configurations for accommodating the $P$ localized holes
on the $N$ available sites.
The lattice potential $ W(u) $ is therefore expected
to have up to
(\raise0.9ex\hbox{$N$}\kern-0.9em\lower1.0ex\hbox{$P$})
nearly degenerate local minima, denoted by $u^{\xi}$
in the following,
corresponding to the 
(\raise0.9ex\hbox{$N$}\kern-0.9em\lower1.0ex\hbox{$P$})
different centroid configurations 
$\xi \equiv ( \xi_1,\dots,\xi_P )$ \cite{Zhong}.
Here, $\xi_i\equiv(\xi_{i,x}\xi_{i,y})$
denotes lattice (centroid) site occupied by the $i$-th hole.
As noted above, each of these local-minimum configurations
breaks the translational symmetry of the lattice
at the level of the 0th order adiabatic approximation.
The symmetry is restored 
in the 1st order adiabatic approximation
by polaron tunneling processes
between the different $u^{\xi}$.

At EP coupling strengths $E_{\rm P}$ larger than, but sufficiently
close to $E_{\rm P}^{({\rm crit})}$, it is possible
that some of the 
(\raise0.9ex\hbox{$N$}\kern-0.9em\lower1.0ex\hbox{$P$})
centroid configurations $\xi$ do not have corresponding
stable local-minimum configurations $u^{\xi}$ in $W(u)$.
This may happen, for example, in a two-hole system ($P=2$),
if one tries to accommodate the
two polarons at 1st neighbor sites, 
$\xi_1$ and $\xi_2$, in the presence of a
1st neighbor Coulomb repulsion $V_{\rm C}$. 
At sufficiently strong $V_{\rm C}$, the corresponding local minimum
$u^\xi\!\equiv\!u^{(\xi_1,\xi_2)}$ 
becomes locally unstable, which is signaled by the
smallest eigenvalue of the restoring force matrix
$\partial^2 W/\partial u^2|_{u^{\xi}}$ becoming negative.
In the following, we will not consider such situations, but rather
restrict ourselves to parameter regions where all
the local minimum configurations
$u^{\xi}$ are stable.

To establish the basic structure of the effective polaron
tunneling dynamics, we treat the path integral
for the effective action $S_{\rm Ad}$ (\ref{eq:s_eff}) 
or its equivalent Hamiltonian $H_{\rm Ad}$
by a lattice dynamical many-body tight-binding approach.
The basic idea behind this approach is that an effective 
polaron tunneling Hamiltonian $H_{\rm P}$ can be defined 
which operates in a ``low-energy'' sub-space of nearly orthogonal 
tight-binding basis states $\mid \phi_\xi \rangle$, 
labeled by the localized polaron centroid configurations $\xi$.
Each such basis state represents a
lattice wavefunction $\phi_\xi(u)$ which
is assumed to be localized in $u$-space around the corresponding 
local potential minimum configuration $u^\xi$. 
For example, $\phi_\xi(u)$
could be chosen as the vibrational (``phonon'') groundstate obtained
in a harmonic approximation by expanding $W(u)$ to quadratic
order around $u^\xi$. 
By restricting the lattice Hilbert space to such a set of basis states
$\phi_\xi$, all vibrational excited states 
around the polaronic local minima
are neglected. Thus, formally, our approach can be regarded as a
tight-binding approximation, formulated for the quantum
dynamics of the multiple-well lattice potential $W(u)$ in the 
$N$-dimensional lattice configuration ($u$-) space.

In the simplest tight-binding approach one would then simply
estimate the matrix elements of $H_{\rm P}$
by projecting the adiabatic lattice
Hamiltonian $H_{\rm Ad}$ onto the corresponding tight-binding 
low-energy sub-space
spanned by all $\phi_\xi$. In such a 1st order projection
approach, one neglects all 
effects arising from virtual excitations 
out of the low-energy sub-space.

Since tunneling matrix elements are exponentially sensitive
to small corrections in, for example, the tunneling
barriers, such a tight-binding projection
could cause severe quantitative errors in the estimation
of the magnitude of tunneling matrix elements.
Also, as a practical matter, the accurate evaluation of 
Hamiltonian matrix elements
with basis functions defined on the $N$-dimensional 
$u$-space can become quite difficult.
Lastly, in the 1st order projection 
approach, it is difficult to include the tunneling 
Berry phases into $H_{\rm P}$. 

To avoid the foregoing difficulties, we have adopted
an approach which is based on a direct mapping
of imaginary-time tunneling paths,
rather than a mapping of Hamiltonian matrix elements.
Formally, this is accomplished by the path integral instanton
method \cite{Rajaraman,Coleman,Yonemitsu}. In this method,
an ``instantaneous'' polaron hopping process $\xi\to\zeta$ induced
by $H_{\rm P}$ between two polaron centroid configurations
$\xi\equiv(\xi_1\dots\xi_P)$ and $\zeta\equiv(\zeta_1\dots\zeta_P)$
is identified with the (restricted) path
sum of instanton tunneling paths connecting
$u^\xi$ to $u^\zeta$ in $u$-space.
The effective action $S_{\rm P}$
of the instantaneous hopping paths, so obtained,
can then be immediately translated into matrix elements
of the effective tunneling Hamiltonian $H_{\rm P}$.
Since only tunneling paths, but no basis states
enter into the mapping, the results do not depend
on any particular choice of tight-binding basis 
states  $\phi_\xi$.

As a specific starting point, we consider the trace of 
the resolvent operator at complex energies $E$
\begin{equation}
{\rm Tr} \ (E-H)^{-1} = -\int_0^\infty d \beta \ 
e^{\beta E} \ {\rm Tr} \ e^{-\beta H} 
\;,\label{eq:resolvent}
\end{equation}
written in the imaginary-time domain in path-integral form, 
\begin{equation}
{\rm Tr} \ e^{-\beta H} = \int_{u(\beta) = u(0)} {\cal D}u(\tau) \ 
e^{- S_{\rm Ad} [u(\tau)]} 
\;.
\label{eq:trace_eb}
\end{equation}
The trace operation in Eq.~(\ref{eq:resolvent}) leads to periodic
boundary conditions on the imaginary time interval 
$[0,\beta]$ in Eq.~(\ref{eq:trace_eb}). 
These periodic boundary conditions
in Eq.~(\ref{eq:trace_eb}) impose not only the closed path constraint
$u(\tau)=u(0)$, but also the condition that the initial
and final electron wavefunctions must be the same, including
their phase factors. That is, for the electron wavefunctions 
$\mid\Psi(u(\tau))\rangle$ entering into $S_{\rm Ad}$ via 
Eq.~(\ref{eq:overlap}), the constraint 
$\langle \Psi(u(\beta)) \mid \Psi(u(0))\rangle=+1$ must be imposed 
for all paths $u(\tau)$ integrated over in Eq.~(\ref{eq:trace_eb}).
The latter requirement ensures that the Berry phase contribution
to $S_{\rm Ad}$ in Eq.~(\ref{eq:trace_eb}) is uniquely defined 
for every closed path $u(\tau)$, independent of the choice of phase 
for each individual electronic wavefunction 
$\mid \Psi(u(\tau))\rangle$ along such a path.
Quantized eigenenergies can be found from
(\ref{eq:resolvent}) by searching 
for the poles of the trace of the resolvent operator on the
real $E$-axis.

The main contributions to the low-energy part of 
(\ref{eq:resolvent}) arise from instanton path
configurations, i.e., $u$-paths which are almost 
always close to one of the centroid configurations, occasionally 
transfer from one to another centroid configuration by an
almost instantaneous polaron 
hopping process, and finally return to the initial 
$u$-configuration at imaginary time $ \beta $,
in order to satisfy the closed path constraint.
Important closed-path tunneling processes for polaron states with 
$P$=1 and 2 dopant-induced holes 
are shown in Fig.~\ref{fig:closed_path}.
Each black circle represents an occupied polaron centroid
site in the initial configuration $\xi$ of the hopping process.
Arrows indicate the hopping processes transferring the
initial configuration $\xi$ into the final configuration $\zeta$.
Thus, in $u$-space each arrow corresponds to a set of instanton-type
tunneling paths which connects the two respective minimum-$W$ endpoint
configurations $u^\xi$ and $u^\zeta$
and traverse the $W$-barrier 
separating the two minima.\cite{Zhong}
Note that, as discussed above,
via such tunneling paths, a hole {\em polaron \/} can 
tunnel in a single step between 2nd-, 3rd-, etc. neighbor 
sites even if the original {\em electron \/} Hamiltonian 
(\ref{eq:model}) contains only a 1st-neighbor $t$.\cite{Zhong}

First, we consider the case of $P$=1. For the time being,
we  take into account 
only the 2nd and 3rd neighbor 
processes denoted by amplitudes $ t_1^{(2)} $ and $ t_1^{(3)} $ in 
Fig.~\ref{fig:hopping}(a).
Single-polaron {\it inter\/}-sublattice processes are strongly 
suppressed by the AF spin correlations.\cite{Trugman} Hence,
the 1st-neighbor amplitude $ t_1^{(1)} $ can be much smaller than 
or, at most, comparable to $ t_1^{(2)} $ and $ t_1^{(3)} $ 
(to within 20-30\%) in the case of $P$=1.
Then, instanton path configurations are classified according to 
the numbers of intra-sublattice processes: 
$ n_x $ counts the number of $ t_1^{(3)} $ processes to the right, 
$ m_x $ the $ t_1^{(3)} $ processes to the left, 
$ n_y $ the $ t_1^{(3)} $ processes to the upper, 
$ m_y $ the $ t_1^{(3)} $ processes to the lower, 
$ n_u $ the $ t_1^{(2)} $ processes to the upper-right, 
$ m_u $ the $ t_1^{(2)} $ processes to the lower-left, 
$ n_v $ the $ t_1^{(2)} $ processes to the lower-right, and 
$ m_v $ the $ t_1^{(2)} $ processes to the upper-left neighbors.
Path integration over the corresponding instanton paths
gives\cite{Rajaraman,Coleman,Yonemitsu}
\begin{eqnarray}
&&{\rm Tr} \ e^{-\beta H}= \nonumber\\
&&e^{-\beta W(u^{({\rm min},1)})} \sum_{n_x,\dots,m_v} 
\frac1{n_x! m_x! n_y! m_y! n_u! m_u! n_v! m_v!} \nonumber \\
&\times & \int \frac{d p_x}{2\pi} 
e^{i p_x(2n_x-2m_x+n_u-m_u+n_v-m_v)} \nonumber \\
&\times & \int \frac{d p_y}{2\pi} 
e^{i p_y(2n_y-2m_y+n_u-m_u-n_v+m_v)} \nonumber \\
&\times & (e^{-\delta R_{\rm 1}^{(2)} -i \theta_1^{(2)}}
J_1^{(2)}K_1^{(2)}\beta)^{ n_u + m_u + n_v + m_v } \nonumber \\
&\times & (e^{-\delta R_{\rm 1}^{(3)} -i \theta_1^{(3)}}
J_1^{(3)}K_1^{(3)}\beta)^{ n_x + m_x + n_y + m_y } \nonumber \\
&=& \int \frac{d p_x}{2\pi} \frac{d p_y}{2\pi} \exp \left[
-\beta \left\{ W(u^{({\rm min},1)}) +2 t_1^{(2)} [ \cos(p_x + p_y) 
\right. \right. \nonumber \\ 
&& 
+ \cos(p_x-p_y) ] \left. \left. +2 t_1^{(3)} 
[ \cos(2p_x) + \cos(2p_y) ] \right\} \right]
\;.\label{eq:one_polaron}
\end{eqnarray}
The effective hopping matrix elements
$ t_P^{(\nu)} $ are obtained as 
\begin{equation}
t_P^{(\nu)} = - J_P^{(\nu)} K_P^{(\nu)}
e^{-\delta R_P^{(\nu)} -i \theta_P^{(\nu)}} 
\;.\label{eq:effective_hopping}
\end{equation}
$ W(u^{({\rm min},1)}) $ is the absolute minimum lattice potential 
energy obtained at a minimum-$W$ configuration 
$ u^{({\rm min},1)}\equiv u^{(\xi_1)} $
for $P=1$.
Factorial factors such as $ n_x! $, etc. are 
introduced to account for identical 
species of instantons. The $ p_x $ and $ p_y $-integrals are 
introduced to enforce the imaginary-time periodic boundary
condition.
The quantity $ \delta R_P^{(\nu)} $ is the 
single-instanton contribution to the real part of the action 
for the path segment of the corresponding
tunneling process $t_P^{(\nu)}$ 
and $\theta_P^{(\nu)}$
is the corresponding Berry phase contribution.
The assignment of a unique Berry phase factor 
$e^{-i \theta_P^{(\nu)}}$ to each such open path segment
requires more detailed symmetry considerations and will
be postponed until Sec.~\ref{sec:symmetry}.
The quantity $ K_P^{(\nu)} $ in (\ref{eq:effective_hopping}) 
represents the $-\frac12$-th power of the fluctuation determinant 
for the instanton solution with the zero mode excluded 
divided by that for the static solution at $ u^{({\rm min},1)} $, 
and $ J_P^{(\nu)} $ is the Jacobian needed for a special 
treatment of the corresponding zero mode.
They are defined as in Eqs.~(10.13) and (10.14) of 
Ref.~\onlinecite{Rajaraman}
for the periodic potential problem.
Substituting the result of the path integral (\ref{eq:one_polaron}) 
into the formula (\ref{eq:resolvent}), we obtain the dispersion 
relation shown in the parenthesis of $\exp \left[ -\beta 
\{ \dots \} \right]$ in Eq.~(\ref{eq:one_polaron}). 
Note that the effective hopping matrix element is 
defined such that it is positive if the corresponding
path segment carries a nontrivial $(-1)$ Berry 
phase factor: the sign convention of our polaron tunneling
matrix elements $t_P^{(\nu)}$ is opposite to that used in the 
original electron Hamiltonian (\ref{eq:model}).

Next, we consider the case of $P=2$.
Since self-localization reduces substantially the polaron kinetic 
energy scale, it is favorable for two polarons in an AF
spin background to be bound in a pair:
the binding energy can easily be of the order of the effective 
polaron nearest-neighbor attraction, i.e., comparable to 
the AF spin exchange coupling $ J $\cite{phase_sep} in the 
Holstein-$ tJ $ model.
As a first approximation, we therefore restrict the
path integration to include only
1st neighbor pair configurations $u^{(\xi_1,\xi_2)}$
and the instanton tunneling paths connecting them.
Our numerical studies described below suggest
that these 1st neighbor configurations represent the
{\it absolute} minimum of $W(u)$ for $P=2$.
Other, more distant pair configurations with $|\xi_1-\xi_2|>1$
are either represented by local $W$-minima $u^{(\xi_1\xi_2)}$
of higher energy or they don't form local minima
in $W(u)$ at all.
We are thus limiting ourselves, for now, to the tunneling
processes $ t_2^{(2)} $ and $ t_2^{(3)} $ between the
degenerate, absolute-minimum $u$-configurations
as shown in Fig.~\ref{fig:hopping}(b).

The technique used above for $P=1$ can be generalized in a 
straightforward manner to the present case $P=2$. Here, in addition
to the lattice translational degeneracy of the minimum-$W$
$u$-configurations, the $P=2$ system exhibits 2-fold
internal degeneracy, corresponding to the two possible
orientations of the 1st neighbor polaron pair, along either
the $x$- or along the $y$-axis. Because of this 2-fold
internal degree of freedom, the instanton exponential
function in the path integral takes the form of the trace
over a $2\times2$ matrix exponential, namely
\begin{eqnarray}
&& {\rm Tr} \ e^{-\beta H} = \nonumber \\
&& \int \frac{d p_x}{2\pi} \frac{d p_y}{2\pi} \ {\rm Tr} \ \exp \left[
-\beta W(u^{({\rm min},2)}) 
\left( \begin{array}{cc} 1 & 0 \\ 0 & 1 \end{array} \right)
\right. \nonumber \\ 
&& \left.
-\beta \left( \begin{array}{cc} 
2 t_2^{(3)} \cos p_x & 
4 t_2^{(2)} \cos \frac{p_x}{2} \cos \frac{p_y}{2} \\ 
4 t_2^{(2)} \cos \frac{p_x}{2} \cos \frac{p_y}{2} & 
2 t_2^{(3)} \cos p_y \end{array} \right) \right]
\;,\label{eq:two_polaron}
\end{eqnarray}
where $ W(u^{({\rm min},2)}) $ denotes 
the absolute minimum lattice potential energy for $P=2$,
obtained at $ u^{({\rm min},2)} \equiv u^{(\xi_1\xi_2)} $ with 
$\xi_1$ and $\xi_2$ denoting 1st neighbor centroid sites.
The tunneling matrix elements
$ t_P^{(\nu)} $ are expressed analogous
to Eq.~(\ref{eq:effective_hopping}) in terms of the
action contributions, fluctuation determinants, and Jacobians 
of the respective instanton path segments.
The 2 low-lying eigenenergies of the polaron pair
at total momentum $p\equiv(p_x,p_y)$
are obtained by diagonalizing the $2\times 2$
matrix in $\exp \left[ -\beta \{ \dots \} \right]$ 
of Eq.~(\ref{eq:two_polaron}). 

The generalization of the foregoing path integral approach
to $P>2$ hole polaron states is in principle straightforward,
but becomes practically difficult to implement
with increasing polaron number $P$. 
Analogous to (\ref{eq:two_polaron}),
the approach leads to a momentum integral over the trace
of a matrix exponential where the matrix dimension reflects
the number of (nearly) degenerate, translationally inequivalent
polaron centroid configurations $(\xi_1\dots\xi_P)$ included
in the tunneling analysis. In the following, we restrict ourselves
to the cases $P=$0, 1, and 2 which will allow us to extract the
effective 1-polaron tunneling and 2-polaron interaction matrix
elements. 

The low-lying tunneling eigenenergies identified by 
the foregoing instanton path integral method 
(and their corresponding eigenstates)
can be equivalently represented in terms of
an effective polaron tunneling Hamiltonian $H_{\rm P}$
where each polaron is represented as a spin-$1\over2$
fermion. $H_{\rm P}$ is thus
defined to operate in an effective spin-${1\over2}$ fermion 
Hilbert space with the effective fermions occupying sites
$\xi_i$ on the 2D square lattice of the original EP Hamiltonian.
A $P$-polaron centroid configuration $(\xi_1\dots\xi_P)$
is mapped onto the corresponding state of $P$
site-localized fermions with minimum possible total spin,
i.e., with $S_{tot}={1\over2}$ (0) for odd (even) $P$.
The latter mapping condition reflects the fact
that the absolute electron
groundstates $\mid\Psi(u)\rangle$, 
numerically calculated on finite clusters,
exhibit minimum total spin quantum  number.
Notice however that by representing the polaron as an effective 
spin-$1\over2$ fermion, we are actually including
low-energy spin excitations into the effective Hamiltonian 
description. In order to derive the effective polaron spin-spin
interactions, our adiabatic path integral
treatment can be straightforwardly generalized
to include restricted electron groundstates in
Hilbert space sectors of higher total spin quantum numbers
$S_{tot}\geq 1$. In this manner, the spin-$1\over2$ fermion 
representation can be extended well beyond the scope of our original 
adiabatic approximation which retains only the (minimum-spin)
absolute electron groundstate $\mid\Psi(u)\rangle$.
In the following analysis, we limit ourselves to the absolute 
groundstate only. Hence we are only studying the
total spin-singlet pair state in the $P=2$ case.
Using our numerical Berry phase results, we will show
in Sec.~\ref{sec:symmetry}
that each single  polaron in such a singlet pair behaves 
indeed as a spin-$1\over2$ fermion.

In generalizing the above 1st neighbor approach,
it is also straightforward to include inter-sublattice hopping 
processes:
the dimension of the matrix increases, the ${\bf k}$-independent 
term is no longer proportional to the unit matrix, and $ t_P^{(1)} $ 
(more precisely, $ t_1^{(1)} $, $ t_2^{(1a)} $, and $ t_2^{(1b)} $) 
are defined as above.
Then, the effective Hamiltonian describing the polaron
tunneling dynamics and effective polaron-polaron
interactions can be written in the form, 
\begin{eqnarray}
H_{\rm P} =&& \sum_{i \neq j, \sigma} 
 ( t_{ij} + \sum_k \Delta t_{ijk} n_{d k})
\nonumber \\
&&\;\;\;\;\times(1-n_{d j,-\sigma})
d_{j \sigma}^\dagger d_{i \sigma}
(1-n_{d i,-\sigma}) 
\nonumber \\
&& -\sum_{\langle i,j \rangle}
V_{\rm P} n_{d i} n_{d j}
\;.\label{eq:effective_Hamiltonian}
\end{eqnarray}
Thus, $d_{j\sigma}^\dagger $ creates a spin-$1\over2$-fermion polaron 
with spin $ \sigma $ at site $ j $, 
$ n_{d j} = \sum_\sigma d_{j\sigma}^\dagger d_{j\sigma} = 0, 1 $ 
and $ P = \sum_j n_{d j} = 1, 2 $.
The hopping term is to include, appropriately, the amplitudes 
$ ( t_{ij} + \sum_k \Delta t_{ijk} n_{dk}) \equiv t_1^{(1)} $, 
$ t_1^{(2)} $, $ t_1^{(3)} $, $ t_2^{(1a)} $, $ t_2^{(1b)} $, 
$ t_2^{(2)} $, or $ t_2^{(3)} $ [with appropriate sign 
according to the corresponding Berry phase factor]
for $ i \rightarrow j $ tunneling processes 
shown in Fig.~\ref{fig:hopping}. Note here that the sign convention
for the polaron tunneling amplitudes $t_P^{(\nu)}$ in
(\ref{eq:effective_Hamiltonian}) is opposite to that used in
the underlying electron Hamiltonians (\ref{eq:h_tj},\ref{eq:h_hub}).

The 1st neighbor attraction $ V_{\rm P} $ in 
(\ref{eq:effective_Hamiltonian}) is estimated as
\begin{equation}
V_{\rm P} = 
2 W(u^{({\rm min},1)}) - W(u^{({\rm min},2)}) - W(u^{({\rm min},0)}) 
\;, \label{eq:effective_attraction}
\end{equation}
where $ W(u^{({\rm min},P)}) $ is the (absolute) 
minimum potential energy $ W(u) $ 
for the $ u $-configurations
$u^{({\rm min},P)}\equiv u^{(\xi_1\dots\xi_P)}$ 
which minimize $W(u)$
for $ P $ holes. For $ P $=2, our numerical calculations
suggest that the absolute minimum-$ W $ $ u $-configuration 
does indeed correspond to the 1st neighbor pair.
For purposes of estimating $V_{\rm P}$ numerically on
small model clusters, we have minimized $W_0$ instead of $W=W_0+W_1$,
thus neglecting the effect of $W_1$ on $u^{({\rm min},P)}$.
In the physical parameter regime of interest, $\Omega\ll t, E_{\rm P}$,
these $W_1$-corrections to $u^{({\rm min},P)}$ are indeed small, 
of order $(\Omega/t)^2$. The full potential $W=W_0+W_1$ was used
to calculate $ W(u^{({\rm min},P)}) $.

To obtain order of magnitude estimates for $ t_P^{(\nu)} $, 
we have used both the dilute-instanton-gas 
approach\cite{Rajaraman,Coleman,Yonemitsu}, as explained above, 
and a constrained lattice dynamics approach\cite{Zhong} which is more 
straightforward and adopted in Sec.~\ref{sec:effective}.
The two approaches have given similar results.
In the latter approach, the lattice Schr\"odinger  equation 
corresponding to $ S_{\rm Ad} $ is solved exactly for $ u $ 
constrained to the linear tunneling path $u^{(\zeta\xi)}(s)$
which is defined analogous to Eq.~(\ref{eq:tun_path})
and connects the two
energetically degenerate, minimum-$ W $ polaron end-point 
$ u $-configurations $u^{\xi}$ and $u^{\zeta}$ of the respective hop.
The hopping matrix element $ |t_P^{(\nu)}| $ is then estimated as 
one half of the ground-to-1st-excited state energy splitting.

\section{SYMMETRY OPERATIONS AND BERRY PHASES}
\label{sec:symmetry}

Before going into numerical estimations of effective model 
parameters, we need to settle the quasiparticle statistics and 
the signs of effective polaron hopping processes by calculating 
Berry phase factors.
To calculate $ \exp( -i \theta [u(\tau)] ) $ for tunneling paths 
$ u(\tau) $ shown in Fig.~\ref{fig:closed_path}, 
we discretize $ \tau $ with at least 5 $ \tau $-points 
per linear path segment and obtain $\mid \Psi(u(\tau)) \rangle$ of the 
Holstein-$ tJ $ model by the Lanczos exact diagonalization 
method on an $ N $=4$ \times $4 lattice with 
periodic boundary conditions.
The electron Hilbert space is restricted to the sector of 
minimum total spin ($S$=0, $1/2$, 0 for $P$=0, 1, 2, respectively), 
which comprises the absolute ground state 
$\mid \Psi(u) \rangle$ for $ u $-configurations 
near the local $ W $-minima.
The results for all paths in Fig.~\ref{fig:closed_path} are 
summarized by 
\begin{equation}
\theta [u(\tau)] = \pi \left( 
m^{(2)} + m^{(3)} + m_2^{(1)} \right)
\;, \label{eq:closed_path}
\end{equation}
where $ m^{(\nu)} $ is the number of $ \nu $-th neighbor hops 
with $ \nu $=2, 3, and $ m_2^{(1)} $ for $ P $=2 denotes 
the number of 1st-neighbor hops indicated by the dashed bonds shown 
in Fig.~\ref{fig:parity}(a) by the first polaron in close 
proximity to the second, static polaron, indicated as a black circle.
The effect of the $ m_2^{(1)} $-term can be illustrated, for example,
by comparing the Berry phase factors $e^{-i\theta}$ of 
the triangular paths (a)(A) and (b)(B) shown in
Fig.~\ref{fig:closed_path}. 
In both paths, a single polaron
is taken around the triangle in 3 steps, consisting of two 
1st neighbor and one 2nd neighbor transfer. For the one-polaron case,
(a)(A), the phase factor is $(-1)$, for the two-polaron case (b)(B) 
it is $(+1)$. Thus, the close proximity of the second,
static polaron in (b)(B) has altered the Berry phase of the first 
polaron tunneling around a closed path.

The origin of the $ m_2^{(1)} $-term can be traced back to the
internal symmetry of  $\mid \Psi(u) \rangle$: 
For the local minimum-$ W $ $ u $-configurations of
2nd- and 3rd-neighbor polaron pairs, $\mid \Psi(u) \rangle$ 
is odd under reflection along the dashed lines shown 
in Fig.~\ref{fig:parity}(b), i.e., along the 
pair axis for the 2nd-neighbor pair and perpendicular to
the pair axis for the 3rd-neighbor pair.
Suppose, for example, that the first polaron hops from (1,0) to 
(1,1) in a first step and from (1,1) to (0,1) in a second step with 
the second polaron staying fixed at (0,0). These are the first 
two steps of path (b)(B) in Fig.~\ref{fig:closed_path}.
Note that the two steps generate the same final centroid configuration
as a reflection along the dashed (0,0)--(1,1) line, shown in
Fig.~\ref{fig:parity}(b).
Because of this odd ``internal'' parity of $\mid \Psi(u) \rangle$
for the intermediate (2nd neighbor polaron pair) configuration, 
one of the two 1st neighbor hops must 
contribute an additional factor $(-1)$. Assigning this
$(-1)$ phase factor to one of the two 1st neighbor steps
in path (b)(B) of Fig.~\ref{fig:closed_path}
is to some extent arbitrary. The pattern
of dashed-line and full-line bonds surrounding the static
polaron in Fig.~\ref{fig:parity}(a) represents one possible
assignment which is consistent with all the closed-path Berry phase
results in Fig.~\ref{fig:closed_path}(b).
As a consequence of its odd internal parity,
the 2nd neighbor polaron pair configuration is actually allowed to 
contribute with finite amplitude to polaron pair wave functions of 
$ d_{x^2-y^2} $ symmetry, in spite of the fact that the 2nd neighbor
pair axis points along 
the nodal axis of the $d_{x^2-y^2}$ pair wavefunction.

The $ m^{(2)} $- and $ m^{(3)} $-terms can be regarded as due to 
strong antiferromagnetic correlation.
Suppose a polaron is initially located at (0,0) and hops to (2,0), 
(1,1), and then back to (0,0) along the path (a)(B) in 
Fig.~\ref{fig:closed_path}. The electron initially located 
at (2,0) hops to (0,0) and then to (1,1), while the electron 
initially located at (1,1) hops to (2,0). Thus, if one approximates 
the AF spin background by a N\'{e}el state, two electrons of like spin
are exchanged. This produces a fermionic $(-1)$-factor.
More generally, when a closed path consists of an odd number of 2nd- 
or 3rd-neighbor hopping processes, an even number of electrons 
within a sublattice are cyclically permuted, producing the $(-1)$ 
factor within the N\'{e}el approximation to the AF spin background.
In order for this to occur, the AF spin correlation has to be strong, 
but it need not be long-ranged.
For $P$=1, the Berry phase rule can be completely explained 
in this way.

For both $P$=1 and 2, $ \theta [u(\tau)] $ is given by a sum of 
independent single-polaron hopping contributions and 
$ \exp( -i \theta [u(\tau)] ) $ does {\em not \/} depend on 
whether or not the two polarons are being 
adiabatically exchanged in a given path 
[Fig.~\ref{fig:closed_path}(b)].\cite{Arovas}
Thus, for example, paths (b)(B) and (b)(C) in 
Fig.~\ref{fig:closed_path} contain the same 
1st and 2nd neighbor hops and they have the same Berry phase.
The two paths differ only in that
(b)(C) exchanges the two polarons, whereas (b)(B) does not.
Since the pair is a total spin singlet, this implies that 
each single polaron in the pair behaves effectively as a 
spin-$1/2$ fermion or as a spin-0 boson.
Only the spin-$1/2$ fermion representation is consistent with 
the half-odd-integer total spin in odd-$ P $ systems and, as discussed
in the previous sections, it is the one we have adopted.
Equation~(\ref{eq:closed_path}) rules out the possibility of 
representing dopant-induced hole polarons as spin-0 fermions 
or as spin-$1/2$ bosons.

To settle the signs of effective polaron hopping processes, we need 
to define Berry phase factors for the corresponding single-hop
open path segments. 
Let the initial $ u $-configuration of such a single-hop
path segment be denoted by $ u^{(\xi)} $ and 
the final $ u $-configuration by $ u^{(\zeta)}$.
The assignment of a Berry phase to such a path
segment can be made unique by fixing the
phase of the corresponding wavefunction 
$\mid\Psi( u^{(\zeta)})\rangle$
relative to that of $\mid \Psi( u^{(\xi)})\rangle$
in some unique manner.
Given $u^{(\xi)}$ and $\mid \Psi(u^{(\xi)})\rangle$, 
let $\mid\Psi^{({\rm ref})}(u^{(\zeta)})\rangle$ denote such a
final-state reference wavefunction.
Also, let $\mid\Psi^{({\rm ad})}(u^{(\zeta)})\rangle$
denote that groundstate wavefunction $\mid \Psi(u^{(\zeta)})\rangle$
which one obtains by adiabatically 
evolving $\mid\Psi(u)\rangle$ along the tunneling path segment,
without discontinuity in phase, beginning with 
$\mid \Psi(u^{(\xi)})\rangle$.
The Berry phase of the path segment 
is then defined as the phase difference between
$\mid\Psi^{({\rm ad})}(u^{(\zeta)})\rangle$
and $\mid\Psi^{({\rm ref})}(u^{(\zeta)})\rangle$, 
that is, as the phase of the wavefunction overlap 
$
\langle\Psi^{({\rm ref})}(u^{(\zeta)})
\mid
\Psi^{({\rm ad})}(u^{(\zeta)})\rangle
$.
If, for example, $ u^{(\xi)} $ and $ u^{(\zeta)} $ are 
related by a lattice symmetry operation,
we can choose $\mid\Psi^{({\rm ref})}(u^{(\zeta)})\rangle$
as the groundstate wavefunction 
$\mid \Psi(u^{(\zeta)})\rangle$
generated by applying to $\mid \Psi(u^{(\xi)})\rangle$ the symmetry 
operation which transforms $u^{(\xi)}$ into $u^{(\zeta)}$. 
If $u^{(\xi)}$ and $u^{(\zeta)}$ 
are related by several different symmetry operations
giving different $\mid\Psi^{({\rm ref})}(u^{(\zeta)})\rangle$,
we need to specify the reference $\mid \Psi(u^{(\zeta)}) \rangle$, 
i.e., which symmetry operation is chosen to generate the reference 
$\mid \Psi(u^{(\zeta)}) \rangle$ from $\mid \Psi(u^{(\xi)}) \rangle$.
There does not always exist such a symmetry operation to relate 
$\mid \Psi(u^{(\zeta)}) \rangle$ and $\mid \Psi(u^{(\xi)}) \rangle$, 
e.g., the 2nd- or 3rd-neighbor pair and the 1st-neighbor one.
Then, we can arbitrarily choose the phase of 
$\mid \Psi^{({\rm ref})}(u^{(\zeta)}) \rangle$.
The Berry phase factor for the corresponding path is also arbitrary.
Figure~\ref{fig:parity}(a) is an example.
If we chose the different phase (i.e., the negative) of 
$\mid \Psi^{({\rm ref})}(u^{(\zeta)}) \rangle$ for the 
2nd (3rd)-neighbor pair, the signs of 
all the $t_2^{(1a)}$ ($t_2^{(1b)}$) processes would be reversed.

For $ P $=1, we first fix, arbitrarily, the phase of
$\mid \Psi(u^{(\xi)}) \rangle$ for 
the centroid configuration  $\xi$=((0,0)).
All the $\mid \Psi^{({\rm ref})}(u^{(\zeta)}) \rangle$ are then 
uniquely defined by either translation or rotation operations.
The Berry phase factors for $ P $=1 are summarized 
in Fig.~\ref{fig:open_path}(a).
Here, the translation $(x,y) \rightarrow (x+a,y+b)$ 
is denoted by $T(a,b)$, and the rotation 
$(x,y) \rightarrow (x\cos\phi-y\sin\phi,x\sin\phi+y\cos\phi)$ 
is denoted by $R(\phi)$.
The left-hand side shows $\mid \Psi^{({\rm ad})}(u^{(\zeta)}) \rangle$.
The right-hand side shows the possible choices of
$\mid \Psi^{({\rm ref})}(u^{(\zeta)}) \rangle$, generated
from the same $\mid \Psi(u^{(\xi)}) \rangle$ by 
the appropriate lattice symmetry operations.
Note that, for 2nd and 3rd neighbor hops, both
translation and rotation operations generate the same
$\mid \Psi^{({\rm ref})}(u^{(\zeta)})\rangle$.
The first three lines of Fig.~\ref{fig:open_path}(a)
give
$ \exp [ -i \theta_1^{(1)} ] $=$+1$, 
$ \exp [ -i \theta_1^{(2)} ] $=$-1$, 
$ \exp [ -i \theta_1^{(3)} ] $=$-1$, 
and thus 
\begin{equation}
 t_1^{(1)} < 0 \;, \ \ t_1^{(2)} > 0 \;, \ \ t_1^{(3)} > 0 \;. 
\end{equation}
Obviously, these results are consistent with 
the relation (\ref{eq:closed_path}).

For $ P $=2, we first fix the phase of $\mid \Psi(u^{(\xi)}) \rangle$ 
for the centroid configuration 
$\xi\equiv( \xi_1,\xi_2 )$=((0,0),(1,0)).
Rotating this $\mid \Psi(u) \rangle$ by angle $\pi/2$, we define the 
$\mid \Psi^{({\rm ref})}(u^{(\zeta)}) \rangle$ 
for $\zeta\equiv(\zeta_1,\zeta_2 )$=((0,0),(0,1)). 
The $\mid \Psi^{({\rm ref})}(u^{(\zeta)}) \rangle$ 
for the other 1st neighbor pair configurations $\zeta$
are defined by applying translation operations 
to either $\mid\Psi(u^{(\xi)})\rangle$ or to its rotated version
$R(\pi/2)\mid\Psi(u^{(\xi)})\rangle$. 
The resulting Berry phase factors for $ P $=2 are summarized 
in Fig.~\ref{fig:open_path}(b).
The first two lines imply that
$ \exp [ -i \theta_2^{(2)} ] $=$-1$, 
$ \exp [ -i \theta_2^{(3)} ] $=$-1$, 
and thus 
\begin{equation}
 t_2^{(2)} > 0 \;, \ \ t_2^{(3)} > 0 \;. 
\end{equation}
The last two lines in Fig.~\ref{fig:open_path}(b) imply,
by comparison to the first two lines, that available
rotations would generate the same 
$\mid \Psi^{({\rm ref})}(u^{(\zeta)}) \rangle$ 
as the translations.
The 1st-neighbor hops $ t_2^{(1a)} $ and $ t_2^{(1b)} $ are 
positive or negative for the processes indicated by the dashed 
and full bonds, respectively, of Fig.~\ref{fig:parity}(a),
as discussed above.

\section{BERRY PHASES AND QUANTUM NUMBERS}
\label{sec:phase}

Using the effective Hamiltonian 
(\ref{eq:effective_Hamiltonian}) with parameters 
$ t_1^{(1)} $, $ t_1^{(2)} $, $ t_1^{(3)} $, 
$ t_2^{(1a)} $, $ t_2^{(1b)} $, $ t_2^{(2)} $, $ t_2^{(3)} $ 
(with signs determined above), and $ V_{\rm P} $,
we can now calculate the low-energy eigenstates
for the $P=1$ and $P=2$ polaron systems.

In the case $ P $=1, the dispersion relation from $H_{\rm P}$ 
is given by 
\begin{eqnarray}
\epsilon_1({\bf p}) & = &
 2 t_1^{(1)} [ \cos p_x + \cos p_y ]
+2 t_1^{(2)} [ \cos(p_x + p_y) \nonumber \\  
                   &+& \cos(p_x - p_y) ]  
+2 t_1^{(3)} [ \cos(2p_x) + \cos(2p_y) ]
\;.\label{eq:one_polaron_dispersion}
\end{eqnarray}
As mentioned in Sec.~\ref{sec:instanton}, on finite lattices, 
$t_1^{(1)}$ is smaller than the 2nd and 3rd neighbor $t$'s 
for $P=1$. As the cluster size increases, the overlap between 
the two minimum-$W$ wavefunctions connected by the 
$ t_1^{(1)} $ process, $\mid \Psi(u^{(\xi)}) \rangle$ and 
$\mid \Psi(u^{(\zeta)}) \rangle$ for a 1st neighbor bond 
$(\xi,\zeta)$, becomes small. Then, the potential energy 
$W(u)$ would develop a higher barrier for the 1st neighbor hop, 
due to $W_1(u)$, so that $t_1^{(1)}$ would become further 
smaller.
Allowing for arbitrary values of $ t_1^{(2)} $ and 
$ t_1^{(3)} $ but $t_1^{(1)}=0$, the one-polaron band-minimum is 
located at momentum ${\bf p} = (\pi/2,\pi/2)$ 
for $ \mid t_1^{(2)} \mid < 2 t_1^{(3)} $, 
at $(\pi,0)$ 
for $ \mid t_1^{(2)} \mid > 2 t_1^{(3)} $ and $ t_1^{(2)}>0 $, 
and at $(0,0)$ and $(\pi,\pi)$ 
for $ \mid t_1^{(2)} \mid > 2 t_1^{(3)} $ and $ t_1^{(2)}<0 $, 
as shown in Fig.~\ref{fig:phase}(a).
For the physically relevant signs implied by the Berry phase 
factors, 
$ t_1^{(2)}, t_1^{(3)} > 0 $, the momentum 
of the one-polaron band-minimum is thus at $(\pi/2,\pi/2)$ or 
$(\pi,0)$ which lies on the Fermi surface of the noninteracting 
nearest-neighbor tight-binding band model at half filling.
The one-polaron bandwidth is given by 
\begin{equation}
B_1 = \left\{ \begin{array}{lll}
4 t_1^{(2)} + 8 t_1^{(3)} & {\rm for} &
0< t_1^{(2)}<2 t_1^{(3)} \;,\nonumber \\
8 t_1^{(2)} &               {\rm for} &
0<2 t_1^{(3)}< t_1^{(2)} \;.
\end{array} \right.
\end{equation}

For the cluster geometries studied here 
($N = \sqrt{8} \times \sqrt{8}$, 
$N = \sqrt{10} \times \sqrt{10}$ and $N = 4 \times 4$) and with 
only nearest-neighbor terms ($t$, $J$) included in the 
original Hamiltonian (\ref{eq:model}), 
certain ``accidental'' symmetries exist which cause
$ t_P^{(2)} = t_P^{(3)} $. As a consequence, 
the band-minimum is at $(\pi/2,\pi/2)$, and the eigenvalues
of the inverse effective mass tensor at this point are
\begin{equation}
(m_1^{-1})_r = 8 t_1^{(3)} + 4 t_1^{(2)} = 12 t_1^{(2,3)}
\;,
\end{equation}
\begin{equation}
(m_1^{-1})_\phi = 8 t_1^{(3)} - 4 t_1^{(2)} = 4 t_1^{(2,3)}
\;,
\end{equation}
where the subscript $r$ is for the (1,1) direction and 
$\phi$ is for the (1,$-1$) direction.
If we include finite and negative $ t_1^{(1)} $ (representing, e.g.,
the $ N \rightarrow \infty $ limit at finite, fixed hole density,
rather than at fixed hole number $P=1$), then
the band-minimum is shifted by $ t_1^{(1)} $ from $(\pi/2,\pi/2)$ 
to some point $(p,p)$ with $p<\pi/2$ which would 
fall on the Fermi surface 
of the noninteracting band model at corresponding filling.

For $ P $=2, we first consider the tightly-bound pair limit, 
$ V_{\rm P} \gg \mid t_2^{(\nu)} \mid $, where we can approximate 
the polaron pair ground state by including only nearest-neighbor 
pair configurations, thus retaining only the $ t_2^{(2)} $ and 
$ t_2^{(3)} $ matrix elements of $H_{\rm P}$.
The pair dispersion relations are then given by 
\begin{eqnarray}
\epsilon_2^\pm({\bf p}) && =  - V_{\rm P} 
+ t_2^{(3)} ( \cos p_x + \cos p_y ) \nonumber \\  
\pm && \left[ 
t_2^{(3)2} ( \cos p_x \! - \! \cos p_y )^2 
+ ( 4 t_2^{(2)} \cos \frac{ p_x }2 \cos \frac{ p_y }2 )^2
\right]^{1/2}
\;. \nonumber \\ 
\label{eq:two_polaron_dispersion}
\end{eqnarray}
Allowing for arbitrary values of $ t_2^{(2)} $ and $ t_2^{(3)} $, 
the pair wave function in the nearest-neighbor pair approximation 
for $ \mid t_2^{(2)} \mid > t_2^{(3)} $ has $ d_{x^2-y^2} $-wave 
symmetry if $ t_2^{(2)} > 0 $, and $ s $-wave symmetry if 
$ t_2^{(2)} < 0 $, and, in either case, total momentum 
${\bf p}=(0,0)$, as shown in Fig.~\ref{fig:phase}(b), at the 
band-minimum.
For $ \mid t_2^{(2)} \mid < t_2^{(3)} $, the pair ground states 
are multiply degenerate: 
the horizontal pair (with pair axis parallel to the $x$ axis) 
with total momentum ${\bf p}=(\pi,p_y)$ 
for arbitrary $\mid p_y \mid \leq \pi$, and 
the vertical pair (with pair axis parallel to the $y$ axis) 
with total momentum ${\bf p}=(p_x,\pi)$ 
for arbitrary $\mid p_x \mid \leq \pi$ all have the same energy.
The two-polaron bandwidth is given by
\begin{equation}
B_2 \equiv 
\max_{\bf p}\epsilon^-_2({\bf p}) -
\min_{\bf p}\epsilon^-_2({\bf p})
= 4 \mid \mid t_2^{(2)} \mid - t_2^{(3)} \mid
\;.
\label{eq:b2}
\end{equation}
If we take account of 1st-, 2nd-, and 3rd-neighbor pair 
configurations, including also the 1st-neighbor hopping terms, 
$ t_2^{(1a)} $ and $ t_2^{(1b)} $, in second-order 
perturbation theory, the initially degenerate energy 
along ${\bf p}=(\pi,p_y)$, 
\begin{equation}
\epsilon^-_2(\pi,p_y) = - V_{\rm P} -2 t_2^{(3)}
\;,
\end{equation}
is lowered by 
\begin{equation}
\delta \epsilon^-_2(\pi,p_y) =
- \frac{4( 2 t_2^{(1a)2} \cos^2\frac{p_y}2 + t_2^{(1b)2} 
)}{V_{\rm P} +2 t_2^{(3)}} \;.
\label{eq:degeneracy_1}
\end{equation}
It is reasonable for the $ t_2^{(1a)} $-term to favor 
$ p_y = 0 $ because the process of 
$(\xi_1,\xi_2)$=$((0,0),(1,0)) \rightarrow ((0,0),(1,1))$ 
and the process of $((0,0),(1,1)) \rightarrow ((0,1),(1,1))$, 
for example, are in phase, as shown in Fig.~\ref{fig:parity}(a).
The ground states are still doubly degenerate:
the horizontal pair with ${\bf p}=(\pi,0)$ and 
the vertical pair with ${\bf p}=(0,\pi)$, which 
would correspond to $ p_x $- and $ p_y $-wave though 
they are total-spin-singlets.\cite{YZS1}
For $ t_2^{(2)}, t_2^{(3)} > 0 $ implied by the Berry phase 
factors, we thus get either $ d_{x^2-y^2} $- or 
$ p_{x(y)} $-pairing symmetry with total momentum 
${\bf p}=(0,0)$ or ${\bf p}=(\pi,0)[(0,\pi)]$, respectively.

The accidental symmetries for our finite cluster geometries 
(in the absence of longer-range
terms in the original Hamiltonian (\ref{eq:model}) studied here)
lead to $ t_2^{(2)} = t_2^{(3)} $, which is exactly 
on the $ d $-$ p $ phase boundary 
where $ B_2 $ vanishes due to a frustration effect.\cite{Trugman}
So, the energy $\epsilon^-_2({\bf p})$ is independent of ${\bf p}$.
If we take account of 1st-, 2nd-, and 3rd-neighbor pair 
configurations again, in second-order perturbation theory, the 
initially degenerate energy on the $ d $-$ p $ phase boundary, 
\begin{equation}
\epsilon^-_2({\bf p}) \mid_{ t_2^{(2)} = t_2^{(3)} } = 
-V_{\rm P} -2 t_2^{(2,3)}
\;,
\end{equation}
is lowered by
\begin{equation}
\delta \epsilon^-_2({\bf p}) \mid_{ t_2^{(2)} = t_2^{(3)} } = 
- \frac{f({\bf p})}{V_{\rm P} +2 t_2^{(2,3)}}
\label{eq:degeneracy_2}
\end{equation}
with 
\begin{eqnarray} 
f({\bf p}) &&= 4 t_2^{(1a)2} (2+\cos p_x+\cos p_y) \nonumber\\
&&+  4 t_2^{(1b)2} \frac{1-\cos p_x \cos p_y}{2+\cos p_x+\cos p_y} 
\;. \label{eq:degeneracy_2a}
\end{eqnarray}
Then the ground state has $ d_{x^2-y^2} $ symmetry with 
${\bf p}=(0,0)$ for $\sqrt{2} \mid t_2^{(1a)} \mid > 
\mid t_2^{(1b)} \mid$ and $ p_{x(y)} $-wave with 
${\bf p}=(\pi,0)$ [$(0,\pi)$] otherwise.
For the $ N $=4$ \times $4 Holstein-$ tJ $ cluster with 
periodic boundary conditions, an accidental symmetry 
leads to $ \mid t_2^{(1a)} \mid = \mid t_2^{(1b)} \mid $ 
and thus $ d_{x^2-y^2} $-pairing symmetry.
It is reasonable for the $ t_2^{(1a)} $-term to favor 
the $ d_{x^2-y^2} $-wave state because the process of 
$(\xi_1,\xi_2)=((0,0),(1,0)) \rightarrow ((0,0),(1,1))$ 
and the process of $((0,0),(1,1)) \rightarrow ((0,0),(0,1))$, 
for example, have opposite signs.
Also, it is reasonable for the $ t_2^{(1b)} $-term to favor 
the $ p_{x} $-wave state because the process of 
$(\xi_1,\xi_2)=((0,0),(1,0)) \rightarrow ((0,0),(2,0))$ 
and the process of $((0,0),(2,0)) \rightarrow ((1,0),(2,0))$, 
for example, have opposite signs, 
as shown in Fig.~\ref{fig:parity}(a).
Once again we note that the 2nd-neighbor polaron pair 
configuration contributes to the polaron pair wave function 
of $ d_{x^2-y^2} $ symmetry.

\section{EFFECTIVE HOPPING AND ATTRACTION}
\label{sec:effective}

We have seen how total momenta and internal symmetries of 
few-hole-polaron states are determined by the signs and 
relative magnitudes of the effective 
polaron tunneling matrix elements.
In this section, we show numerical estimates of them with 
effective polaron nearest-neighbor attraction and effective 
pair binding energy to see the energy scales of polaron dynamics.
The relative energy scale of kinetic energy to interaction 
strength is controlled by the phonon frequency in 
the original Hamiltonian (\ref{eq:model}).
It is noted again that we use a constrained lattice dynamics 
approach and exactly solve the lattice Schr\"odinger  equation 
corresponding to the effective action (\ref{eq:s_eff}) 
for the lattice-displacement configurations constrained to 
the linear tunneling path of the respective hop, 
as described at the end of Sec.~\ref{sec:instanton}.
The effective lattice potentials are based on Lanczos calculations 
on finite clusters with periodic boundary conditions.
The numerical results should be regarded as very rough 
order-of-magnitude estimates only.
The nearest-neighbor attraction $ V_{\rm P} $ is calculated 
according to the formula (\ref{eq:effective_attraction}).
The pair binding energy $ \Delta $ is estimated in the 
nearest-neighbor pair approximation, according to 
\begin{equation}
\Delta = 2 \epsilon_1({\bf p}_1^{\rm (min)}) - 
           \epsilon_2^-({\bf p}_2^{\rm (min)})
\;,
\label{eq:delta_1}
\end{equation}
where $ \epsilon_1({\bf p}) $ and $ \epsilon_2^-({\bf p}) $ are 
defined in Eq.~(\ref{eq:one_polaron_dispersion}) (with 
$ t_1^{(1)} $=0) and Eq.~(\ref{eq:two_polaron_dispersion}),
respectively, measured relative to the $P=0$ groundstate energy, and 
$ {\bf p}_P^{\rm (min)} $ are the respective (thus different) momenta 
at the band minima discussed above. 
Note that the sign of $\Delta$ is so defined that 
$\Delta>0$ signifies a net attraction, $\Delta<0$ repulsion.

Figure~\ref{fig:parameter}(a) shows the logarithm of the 
dominant 2nd- and 3rd-neighbor hopping amplitudes 
$ t_P^{(2)} $ and $ t_P^{(3)} $ for $ P $=1, 2 
in the Holstein-$ tJ $ model 
on an $ N = \sqrt{8} \times \sqrt{8} $ cluster.
As expected in a polaronic system,\cite{Holstein} all 
$ t_P^{(\nu)} $ are suppressed, roughly exponentially, 
with increasing $ E_{\rm P}/\Omega $ and strongly reduced 
compared to the bare electronic $ t $.
However, for $ E_{\rm P} $ near $ E_{\rm P}^{\rm (crit)} $, 
the $ t_P^{(\nu)} $ can become comparable 
to the phonon energy scale $ \Omega $.
For $ P $=2, the proximity of the second, static polaron 
strongly enhances the amplitudes $ t_2^{(2)} $ and $ t_2^{(3)} $ 
relative to $ t_1^{(2)} $ and $ t_1^{(3)} $.
It is worth noting that this effect occurs 
only in the presence of 
strong electron correlations where bipolaron
formation is prevented by the strong on-site Coulomb
repulsion. By contrast, this effect 
never occurs in ordinary polaronic systems with the 
electron-phonon interaction $E_{\rm P}$ larger than the local
Coulomb repulsion. In the latter case small bipolarons will 
form \cite{AlRa} which are much heavier than polarons. To generate
the above-described delocalization 
(and hence mobility !) enhancement effect,
it is essential to keep the two polarons spatially separated
by strong enough on-site Coulomb effects.

The accidental symmetries leading to $ t_P^{(2)} = t_P^{(3)} $ 
will be lifted on larger lattices and, more importantly, by 
inclusion of longer-range couplings, such as 2nd-neighbor 
hopping $ t' $, Eq.~(\ref{eq:h_t2}), and extended Coulomb repulsion 
$ V_{\rm C} $, Eq.~(\ref{eq:h_lc}),
in the original EP Hamiltonian (\ref{eq:model}),
as will be shown below. 
Due to the exponential dependence of the delocalization
matrix elements $ t_P^{(\nu)} $ 
on the lattice potential parameters, such additional couplings 
can substantially affect the magnitudes of the $ t_P^{(\nu)} $ 
parameters, without necessarily altering the Berry phase factors 
or the predominance of the 2nd- and 3rd-neighbor hopping terms
($t_P^{(2,3)}>t_P^{(1)}$)
and their two-polaron enhancement ($t_2^{(\nu)}>t_1^{(\nu)}$).
The Berry phase factors should be a robust feature
of our model, since they reflect the topological properties of
the relevant tunneling paths relative to certain singular manifolds
of the lattice action in $u$-space.

Figure~\ref{fig:parameter}(b) shows the nearest-neighbor attraction 
$ V_{\rm P} $ and the pair binding energy $ \Delta $, where 
the latter quantity is given by 
\begin{equation} 
\Delta = V_{\rm P} +2 t_2^{(2,3)} -8 t_1^{(2,3)} 
\label{eq:delta_2}
\end{equation}
for $t_P^{(2)}=t_P^{(3)}$, using (\ref{eq:delta_1}).
Since two self-localized nearest-neighbor holes mutually inhibit 
their delocalization, the $ t $-term in the original Hamiltonian 
(\ref{eq:model}) gives a repulsive contribution to $ V_{\rm P} $:
in the parameter range shown in the figure, 
$ V_{\rm P} < 0.342J $ ($ 0.316J $) is substantially reduced 
compared to $ V_{\rm P}(t=0) = 1.00J $ ($ 0.926J $) on 
$ N = \sqrt{8} \times \sqrt{8} $ ($ \sqrt{10} \times \sqrt{10}$) 
sites in the $ t \rightarrow 0 $ limit.
Compared to the $ tJ $ model, $ V_{\rm P} $ can be larger or 
smaller: self-localization reduces the effective polaron hopping 
processes, giving an attractive contribution, and it is more 
effective in the one-hole state than in the two-hole state, 
giving a repulsive contribution.
The binding energy $ \Delta $ is enhanced by the two-polaron hopping 
amplitudes $ t_2^{(2,3)} $, but it is smaller, in most of the 
parameter range shown in the figure, than $ V_{\rm P} $ 
due to the restricted hopping processes for the polaron pair and 
due to the non-negligible $ t_1^{(2,3)} $-term for large $ \Omega $.
In a more realistic theory, the possible competition between 
polaron pairing and phase separation\cite{phase_sep} would need to be 
considered for finite density of holes.

In order to see a finite size effect, we have calculated 
the effective model parameters 
on $ N = \sqrt{10} \times \sqrt{10} $ sites (not shown) 
to compare with those on $ N = \sqrt{8} \times \sqrt{8}$ sites above.
We find no qualitative difference between them.
In the parameter range shown in the figure, the values of 
$ t_P^{(\nu)} $ are different by a factor of two at most, but 
these values are rough order-of-magnitude estimates in any case.
The values of $ V_{\rm P} $ for $ N = \sqrt{10} \times \sqrt{10} $ 
are smaller by a factor of 0.8-0.9.

For the Holstein-Hubbard model, we find results 
(Fig.~\ref{fig:Hubbard}) quite similar to those shown above.
However, the values of $ V_{\rm P} $ are only 30\% of those 
in the Holstein-$ tJ $ model, which are reminiscent of the fact 
that the hole binding energy is larger for the $ tJ $ model 
than for the Hubbard model.
Furthermore, the values of $ t_2^{(2,3)} $ are smaller than 
those of the Holstein-$ tJ $ model for $ \Omega < 0.2t $, 
and the values of $ t_1^{(2,3)} $ are larger by a factor of 
1.6-3.7 in the parameter range shown in the figure.
All these results make the pair binding energy smaller 
in the Holstein-Hubbard model.
For large $ \Omega $, the polaron pair becomes unbound, 
though our results are based on the adiabatic approximation 
and the nearest-neighbor pair approximation so that they 
are less reliable for large $ \Omega $.

We now turn to the effects of 2nd neighbor
electron hybridization and
long-range Coulomb couplings which lift 
the accidental finite-cluster degeneracy, 
$ t_P^{(2)} = t_P^{(3)} $, and thus shift the system 
off the $ d $-$ p $ phase boundary for $ P $=2, already 
in the absence of $ t_2^{(1a,1b)} $-processes.
The 2nd-neighbor electron hopping term 
in the original Hamiltonian (\ref{eq:model}) 
enhances the 2nd-neighbor hopping $ t_2^{(2)} $, 
lowers the 3rd-neighbor one $ t_2^{(3)} $, and thus 
favors the $ d_{x^2-y^2} $-wave symmetry if $ t' $ is positive 
by the definition in Sec.~\ref{sec:model} 
(Fig.~\ref{fig:2nd_n_hopping_plus}), and the effects are 
opposite if $ t' $ is negative 
(Fig.~\ref{fig:2nd_n_hopping_minus}).
Note that, in the noninteracting tight-binding model, 
the positive $ t' $ raises the energy of $p$=($\pi$,0) state 
[thus the energy of $p$=($\pi/2$,$\pi/2$) state is relatively 
lowered] and makes the Fermi surface convex.
A $t'$-term which helps the 2nd-neighbor electron hopping 
also helps the 2nd-neighbor polaron hopping.

The long-range repulsion term enhances $ t_2^{(3)} $ more than 
$ t_2^{(2)} $, so that it favors the $ p_{x(y)} $-wave symmetry 
(Fig.~\ref{fig:long_range_Coulomb}).
This can be understood if we recall the second-order perturbation 
theory with respect to $ t_2^{(1a,1b)} / V_{\rm P} $.
The $ V_{\rm C} $-term raises the energy of the intermediate 
2nd-neighbor pair favoring the $ d_{x^2-y^2} $-wave symmetry, 
compared to that of the intermediate 3rd-neighbor pair favoring 
the $ p_{x(y)} $-wave symmetry.
Note that $ V_{\rm C} $ enhances both of the $ t_2^{(2)} $ and 
$ t_2^{(3)} $ processes.
This happens because the lattice distortion and thus the localization 
potential is weakened by $ V_{\rm C} $.
If $V_{\rm C}$ is too strong, however, it may overcome 
the nearest-neighbor attraction and the polaron pairing 
will then be suppressed altogether. This will be discussed
further in the next section.

\section{POLARON LIQUIDS AND THE CUPRATE SUPERCONDUCTORS}
\label{sec:p_liquid}

To the extent that the qualitative features 
of the above-discussed effective Hamiltonian 
(\ref{eq:effective_Hamiltonian})
and the resulting tunneling and pairing dynamics
remain intact at finite hole doping concentrations,
the foregoing results have some potentially
interesting consequences for the physical
properties of the polaron liquid formed 
at finite polaron densities. In the present section, 
we will speculate on some of these properties
and compare to experimental observations
in the cuprate high-$T_c$ superconductors.\cite{YZS}

If the above-discussed polaron pair state
remains stable and delocalized at finite hole doping, then
formation of a superconducting
polaron pair condensate can occur
at low enough temperatures.
The foregoing discussion has focussed primarily
on the tightly-bound-pair limit where such a condensate
would be formed via Bose condensation of the {\it pre-existing}
polaron pairs. However, the qualitative $ \Omega $-dependences of
the delocalization energies $ t_P^{(\nu)} $ and 
of the pairing potential $V_{\rm P}$, 
shown in Fig.~\ref{fig:parameter}
suggest that with increasing $ \Omega $ (and fixed electronic 
parameters $ t $, $ J $ and $ E_{\rm P} $), such a condensate 
may exhibit a crossover from
tightly-bound-pair to a BCS-like, 
extended-pair behavior: For small $\Omega$, the delocalization
matrix elements $t_2^{(\nu)}$ and resulting
polaron pair bandwidth become small compared 
to the pairing potential $V_{\rm P}$.
Hence tightly bound pairs will form, 
as described above, with a pair wavefunction
extending only over 1-2 lattice constants.
For large $\Omega$ on the other hand, the
polaron bandwidths ($B_1$ and $B_2$)
can become comparable or larger than the pairing
potential, thus leading to a BCS-like extended pair state,
with a pair wavefunction extending over several/many lattice constants.

In the tightly-bound-pair regime, the Bose condensation
$ T_c $ is controlled by the pair density $x_{\rm pair}$
and the pair bandwidth $B_2$, that is, roughly
\begin{equation}
T_c\sim x_{\rm pair} B_2
\label{eq:t_c}
\end{equation}
where $B_2$ is the pair bandwidth (\ref{eq:b2}) and
\begin{equation}
x_{\rm pair}={1\over2}(1-\langle n\rangle) = {1\over2} x
\label{eq:x_pair}
\end{equation}
is the pair concentration, i.e., 
the half of the hole concentration $x$.
$B_2(\Omega)$ and hence $T_c$ decreases
with decreasing $\Omega$. 

In the BCS-like extended-pair regime, 
$T_c$ is controlled by the pair binding energy
$ \Delta $ which decreases
with increasing delocalization energy and hence with increasing
$\Omega$. As a consequence, there must
exist, somewhere in the cross-over regime between
the tightly-bound-pair and the BCS (extended-pair) limits,
an optimal phonon frequency $ \Omega_o $ where
the transition temperature $ T_c(\Omega) $ 
is maximized. $\Omega_o$ is roughly
determined by the condition
\begin{equation}
B_2(\Omega_o) \sim V_{\rm P}
\;.
\label{eq:omega_o}
\end{equation}
and the maximum possible $T_c$ (as a function of $\Omega$ !), 
estimated by extrapolation of (\ref{eq:t_c})
from the tightly-bound-pair side, is of order
\begin{equation}
T_{co}\equiv T_c(\Omega_o)
\sim x_{\rm pair} B_2(\Omega_o) 
\sim x_{\rm pair} V_{\rm P}
\;,
\label{eq:t_co}
\end{equation}
where $B_2(\Omega)$ is the polaron pair bandwidth, 
Eq.~(\ref{eq:b2}),
as a function of phonon frequency $\Omega$.

One crucial, experimentally observable difference 
between the tightly-bound- and 
the extended-pair condensate is the relation between pair
formation and superconducting transition: In the tightly-bound-pair
regime the pairs, and hence the pairing gap $\Delta$
in the excitation spectrum,
can be {\it preformed}. 
That is, the polaron pairs and the
energy gap for pair breaking exist
already at temperatures $T\sim\Delta$ which could
be well above $T_c$, provided that
$\Delta\gg T_c$. By contrast, in the extended-pair BCS-like regime
we expect the pair formation to coincide with the superconducting
transition, that is, the pairing gap should be observable only
at temperatures $T$ below $T_c$ and should vanish at $T_c$.

The existence of such an optimum phonon frequency
implies that $T_c$ exhibits a vanishing
isotope exponent $\alpha$ when $\Omega=\Omega_o$. To show this,
we note that the isotopic mass dependence enters
into the theory only via the phonon frequency $\Omega$, if the
electron-phonon Hamiltonian is parametrized, 
as in (\ref{eq:e_pol}) and (\ref{eq:omega}),
in terms of $E_{\rm P}$ and $\Omega$, since electron-phonon 
potential constants ($C$) and
harmonic restoring force constants ($K$), and hence $E_{\rm P}$,
are of purely electronic origin, i.e., do {\it not}
depend on atomic/isotope masses. Using $\Omega\propto M^{-1/2}$,
from (\ref{eq:omega}), we conclude that
\begin{equation}
\alpha \equiv -{\partial \log T_c \over \partial \log M}
              \Big|_{\rm el}
       ={1\over2}{\partial \log T_c \over \partial \log \Omega}
              \Big|_{\rm el}
\end{equation}
which vanishes at the $T_c$-maximum
$ 
\Omega\!=\!\Omega_o
$. 
The notation $\dots|_{\rm el}$ here means that the derivatives are
to be taken with all purely electronic model parameters ($t$, $U$, $J$,
$E_{\rm P}$, etc.) held constant. The $T_c$-maximum at $\Omega_o$ 
also implies that $\alpha$
is positive in the tightly-bound-pair regime $\Omega_o>\Omega$, 
but negative in the extended-pair regime $\Omega_o<\Omega$.
The vanishing of $\alpha$ at 
$\Omega=\Omega_o$ does however {\it not} imply that $\alpha$
is generically a small number. Quite to the contrary,
because of the strong $\Omega$-dependence of the 
polaron bandwidth parameters, 
we should expect $\alpha$ to attain quite substantial magnitudes,
with $|\alpha|\sim{\cal O}(1)$, as the system is tuned away
from the optimal phonon frequency, i.e., when $\Omega\neq\Omega_o$.

It is tempting to compare the foregoing features of a finite-density
polaron liquid to the observed properties of the cuprates.
The doping-dependence of the superconducting and normal-state
properties of the cuprates is, in some respects, very much reminiscent
of a cross-over from tightly-bound-pair to BCS/extended-pair
behavior: In the underdoped cuprates, there is now a substantial
body of evidence suggesting that the superconducting gap is
pre-existing, in the form of a ``pseudo-gap'', at temperatures
well above $T_c$.\cite{pseudo_gap_exp}
With increasing hole doping concentration $x$,
$T_c$ approaches a maximum, while the pseudo-gap above $T_c$ is 
gradually suppressed, and, in close proximity to the optimal doping 
concentration $x_o$, the pseudo-gap above $T_c$ vanishes.
Well inside the overdoped regime $x>x_o$, there is no detectable
pseudo-gap and $T_c$ rapidly decreases with increasing $x$.

The isotope exponents $\alpha$ in the underdoped cuprates
are typically quite large in magnitude, of order of the classical
$BCS$-value $\alpha_{\rm BCS}={1\over2}$ or larger. However, in 
contrast to conventional BCS-type phonon-mediated superconductors, 
$\alpha$ can be very sensitive to changes in doping and other 
system properties such as impurity concentration and crystal
structure. With increasing hole-doping concentration, 
the observed, usually positive oxygen isotope exponent $\alpha$ 
decreases and becomes very small, typically $<0.05$, at the optimal 
doping concentration $x_o$.\cite{isotope_exp}
It is presently not clear whether $\alpha$
changes its sign for $x>x_o$. Negative $\alpha$-values have been
observed in copper isotope substitutions on less than optimally doped
cuprate materials.\cite{Franck_Cu}

In comparing these experimental results to the foregoing
theoretical picture of a polaron liquid, it is
important to note that, experimentally, the $T_c$-maximum
and the surmised cross-over from tightly-bound-pair
to extended-pair BCS-like behavior is observed as a function of doping
concentration $x$, whereas, in our above theoretical considerations,
we have discussed the cross-over as a function of phonon frequency
$\Omega$. To see how such a cross-over could arise in our
model as a function of doping, we need to consider
the doping dependence of the polaron delocalization matrix
elements $t_P^{(\nu)}$.

As indicated in 
Figs.~\ref{fig:parameter}-~\ref{fig:long_range_Coulomb}, 
the polaron delocalization
matrix elements, and hence the polaron pair bandwidth $B_2$
are rapidly increasing functions of $\Omega$.
At finite doping, these delocalization matrix elements 
will also become dependent on the hole doping concentration 
$x=1-\langle n\rangle$ by the following mechanism: As the polaron
density increases, the localized wavefunctions of nearby holes will
begin to overlap and the holes will begin to mutually
screen out each other's tunneling barriers. This effect can be 
clearly seen in comparing the 1- and 2-hole results 
in Fig.~\ref{fig:parameter}.
For $P=2$, the mere proximity of the second, static polaron 
strongly enhances the tunneling matrix element of the first,
moving polaron, hence $t_2^{(\nu)} > t_1^{(\nu)}$ for $\nu=2,3$. 
Treated at the mean-field level, at finite polaron density, this 
tunneling enhancement effect will cause the (mean-field average)
tunneling matrix elements to increase with the hole doping 
concentration. Thus the effective polaron 
pair bandwidth $B_2=B_2(\Omega,x)$ 
becomes a strongly increasing function of 
of the hole doping concentration $x$. 

According to the cross-over criterion (\ref{eq:omega_o}) 
it may then be possible to drive the
polaron liquid from the tightly-bound-pair into the extended-pair
regime by changing either the phonon frequency $\Omega$ 
or the doping concentration $x$, if $V_{\rm P}$ is
only weakly dependent on
$\Omega$ and $x$. Another way of stating the same result is
to say that the optimal phonon frequency $\Omega_o=\Omega_o(x)$,
according to (\ref{eq:omega_o}), is a decreasing 
function of the hole doping concentration $x$.
The underdoped region corresponds to the 
tightly-bound pre-existing-pair
regime in this picture; the overdoped region is identified with
the extended-pair BCS-like regime.
The superconducting transition  temperature $T_c$
as a function of $x$ reaches a maximum at an optimal doping
concentration $x_o$ not too far from the concentration $x_\Omega$ 
where $\Omega_o(x_\Omega)=\Omega$ and the isotope exponent vanishes.
Notice here that the point $x_o=x_o(\Omega)$ 
[where $T_c(\Omega,x)$ reaches its
maximum as a function of $x$ at fixed $\Omega$] need not
exactly coincide with the point $x_\Omega$ [where the optimal
phonon frequency $\Omega_o(x)$ equals the actual 
phonon frequency $\Omega$].

At sufficiently large hole doping the polaron-polaron wavefunction
overlap and the mutual screening of the hole-localizing potential 
wells $Cu$ may become so strong that the holes become unbound,
that is, the polarons become unstable towards
forming free carriers. This finite-density polaron unbinding
can be regarded as analogous to the Mott delocalization
transition in moderately doped semi-conductors.
The primary difference is that the Mott transition 
in semi-conductors involves the
screening of localizing potential wells due to static impurities
whereas, in the present case, the localizing potential wells
are due to local lattice distortions which are induced,
via the EP coupling, by the polaronic holes themselves.
In the adiabatic potential $W(u)$, this unbinding
will manifest itself in the (gradual or abrupt) disappearance
of local minimum configurations $u^{(\xi)}$. Whether, in the 
thermodynamic limit, this occurs as a sharp transition 
or as a continuous cross-over is presently unclear and 
needs further study \cite{Roeder}.
The nature of the polaron unbinding and the 
characteristic concentration $x_u$ where the unbinding
occurs will also be influenced by the
long-range Coulomb interaction $V_{\rm C}$ and, in more general
EP models, by the spatial range of the EP interaction.\cite{EmHo}

If the optimal polaronic doping concentrations $x_o$ and $x_\Omega$
are close to the polaron unbinding concentration $x_u$, 
the polaron unbinding will likely dominate the cross-over into the
extended pair regime: In this scenario ($x_u\cong x_o,x_\Omega)$,
the cross-over from under- to overdoping
takes the system directly from the tightly-bound polaron pair
liquid into a BCS-like superconductor of extended pairs
of non-polaronic carriers. The effective mass of the non-polaronic 
carriers in the overdoped regime is much less enhanced by the
electron-phonon coupling and, more importantly, the
mass enhancement is independent of the isotopic mass of the ions. 
The latter is suggested by the conventional
weak-coupling electron-phonon theory where the mass enhancement factor
is given by $(1+\lambda_z)$ and the Eliashberg parameter
$\lambda_z$ is independent of the isotope mass.
If the pairing attraction is of predominantly
electronic (i.e., non-phonon) origin, one will then obtain a very
small isotope exponent $|\alpha|\ll 1$ throughout the overdoped regime
$x>x_u$. 

Thus, the overall magnitude of $\alpha$ can 
serve as a distinguishing feature between extended pairs
of polaronic and non-polaronic carriers in the overdoped regime. 
In the former scenario, already described above, $\alpha(x)$ changes 
sign near optimal doping, but $|\alpha|$ well inside the 
overdoped regime can become as large 
as in the underdoped regime, reflecting the fact
that the underlying pair constituents are still single-hole polarons.
By contrast, in the latter (unbound carrier) scenario,
$|\alpha|$ becomes small in the overdoped regime, 
without necessarily incurring a sign change in $\alpha$,
reflecting the non-polaronic nature of the pair constituents.
Further experimental studies of the isotope exponent in the
overdoped cuprate systems would be desirable.

The foregoing features of the underdoped polaron liquid model
and its cross-over into the overdoped regime exhibit strong
similarities with the observed pairing symmetry and 
doping dependences of $T_c$, isotope exponent and pseudo-gap 
in the cuprates. However, in its present form, the model
also suffers from several potential drawbacks which
arise from the {\it small}-polaron character of the self-localized
hole. Small-polaron formation
necessarily implies bandwidths $B_1$ and $B_2$ 
which cannot be substantially larger than the
phonon energy scale $\Omega$, as shown in 
Figs.~\ref{fig:parameter}-~\ref{fig:long_range_Coulomb}. 
As a consequence, small-polaron carriers may be
easily localized by disorder and/or long-range Coulomb interaction 
effects. Also, by Eq.~(\ref{eq:t_co}), the overall magnitude
of the optimal $T_{co}\sim {1\over 2}x B_2\lesssim {1\over 2}x\Omega$ 
cannot exceed some fraction of $\Omega$. With $x\sim 0.10-0.20$
and $\Omega\lesssim 1000 K$\cite{phonon_exp}, 
this upper limit on $T_{c}$ is of
order $50-100 K$ in the cuprates and it is reached
if $E_{\rm P}$ just barely exceeds $E_{\rm P}^{({\rm crit})}$. 
For substantially larger $E_{\rm P}$, $B_2$ and $T_{c}$ are rapidly 
(exponentially !) suppressed with $E_{\rm P}$. It is not clear from 
the experimental data, whether observed carrier mobilities, 
effective masses and $T_c$'s in the underdoped cuprates actually 
exhibit such a strong sensitivity to changes in $E_{\rm P}$
and/or to disorder or long-range Coulomb interactions.

The foregoing limitations of the small-polaron system can
ultimately be traced back to the short-range nature of the 
assumed Holstein EP coupling in our model. Scaling arguments
show that, at the level of the 0th order adiabatic approximation
in spatial dimensions $D\geq2$, short-range
EP models are subject to a dichotomy whereby single carriers either
form small polarons, if $E_{\rm P}$ exceeds a certain threshold
$E_{\rm P}^{({\rm crit})}>0$, or they do not form polarons at all, if
$E_{\rm P}<E_{\rm P}^{({\rm crit})}$ \cite{emin_holstein_scaling}.
By contrast, in systems with additional
longer-range EP couplings, such as 
the Fr\"ohlich model\cite{emin_holstein_scaling,emin_large_pol}, 
as well as 
in 1D short-range EP models\cite{ScHo,turkevich},
it is possible to form large polarons at arbitrarily weak
$E_{\rm P}$, i.e., with $E_{\rm P}^{({\rm crit})}=0$.
It has been argued \cite{emin_large_pol} that large-polaron 
and large-bipolaron models can remedy some of the above described
deficiencies of the small-polaron picture, while retaining
most of the desirable physical features described above.
Thus, in a large-polaron model, there is still preformed pair 
formation above the superconducting $T_c$ in the underdoped regime; 
and the carrier bandwidth is still strongly reduced and dependent 
on isotope mass.
Also, the possibility of a cross-over to a BCS-type free carrier
superconductor, as a function of doping, via polaron unbinding,
is retained in a large-polaron theory.
However the dependence of the bandwidth and $T_c$ on EP coupling 
strengths $E_{\rm P}$ and on phonon frequencies $\Omega$ is only 
algebraic, rather than exponential, and the overall magnitude of 
the large-polaron bandwidths can become substantially larger
than in a small-polaron model, thus allowing for larger
$T_c$'s. The proposed large-polaron theories 
studied so far \cite{emin_large_pol}
have been based on phenomenological continuum models which of course
cannot reproduce lattice-related features, such as the location
of band minima and pairing symmetries discussed above for our
2D lattice model. It will therefore be of interest
to extend our present work to lattice models with longer-range
EP couplings. Such future studies should explore 
the possibility of large-polaron formation and the bandstructure 
and pair wavefunction symmetry of large-polaron pairs.

Another critical problem in the above described polaron models
is the inclusion of long-range Coulomb effects.
Rough estimates based on a point charge model and measured
long-wavelength dielectric constants \cite{SGEH} suggest that
$V_{\rm C}/t$ in the cuprates could be as large as $1-2$,
if only the electronic contribution to the dielectric screening,
that is, only $\epsilon_\infty$, 
is taken into account. If additional screening from phonons,
i.e., $\epsilon_0$, is included, the estimated 
$V_{\rm C}/t$ is reduced to $0.15-0.3$.
The former, $V_{\rm C}/t\sim 1-2$, would be sufficient to
completely suppress the polaron pairing attraction in a system 
containing only two isolated holes, that is, in the limit of vanishing
polaron density. The latter, $V_{\rm C}/t\sim0.15-0.3$,
may be overcome by the AF-mediated 1st neighbor attraction,
but the net attraction strength would still be substantially reduced
by $V_{\rm C}$.\cite{phase_sep,tj_model} 
The suppression of extended pairing states,
such as $d_{x^2-y^2}$ pairing, by the long-range part of the Coulomb
interaction is a common problem in all extended pairing models
which are currently under investigation. Recent studies of
the metallic (in addition to insulating dielectric !) screening
of the extended Coulomb potential at finite
doping density\cite{SGEH,esirgen_et_al}
have suggested that the screened Coulomb potential 
becomes substantially reduced, or even attractive, 
at doping concentrations $x\sim 0.1-0.2$. 
However the foregoing studies are based
on weak coupling or diagrammatic approaches which do not include
polaronic strong-EP effects. It therefore remains
to be seen whether metallic screening in a finite-density
polaron liquid will be sufficient to ``rescue'' the AF-driven
pairing attraction from the repulsive long-range Coulomb forces.

It is also worth re-emphasizing \cite{emin_large_pol}
the strong phonon contribution to the dielectric screening
in the cuprates,
as evidenced by the large measured dielectric constant ratio 
$\epsilon_0/\epsilon_\infty\gtrsim 6$\cite{emin_large_pol,SGEH}.
This phonon contribution, which acts to reduce long-range
Coulomb forces, can be equivalently regarded as a 
long-range attraction, mediated by long-range
(dipolar) EP interactions. This long-range EP interaction
is in fact a primary agent causing
(bi-)polaron formation in the above-cited
\cite{emin_large_pol} phenomenological
large-polaron models. It is therefore quite conceivable
that, in a realistic model of the cuprates, both AF-
{\it and} EP-mediated attractions contribute to the overall
pairing potential and that the EP contribution may even be the
predominant one.

\section{SUMMARY}
\label{sec:summary}

In conclusion, we have developed a treatment of polaron
tunneling dynamics on the basis of a path integral
formulation of the adiabatic approximation.
The adiabatic treatment of polaron tunneling 
has been tested by comparison to exact numerical results
for a two-site Holstein system. The break-down of the
adiabatic approach in the
anti-adiabatic regime has been discussed and the resulting
limitations of applicability for
long-range polaron tunneling processes in lattice models
have been identified.
Using a combination of path integral,
many-body tight-binding and exact diagonalization 
techniques, we have then explored the Berry phases and effective 
matrix elements for single- and two-polaron tunneling, the two-polaron 
quasiparticle statistics, effective two-polaron interactions, 
and polaron pairing states in the 2D Holstein-$ tJ $ and 
Holstein-Hubbard models near half filling.
The effect of 2nd neighbor electron hybridization
and long-range Coulomb repulsion has also been studied.
Due to the AF spin correlations, single-polaron hopping is 
dominated by {\em intra\/}-sublattice 2nd- and 3rd-neighbor 
processes.
These processes are strongly enhanced by close proximity of 
a second polaron.
The Berry phases imply either $ d_{x^2-y^2} $- or 
$ p_{x(y)} $-wave pair symmetries and effective spin-1/2-fermion 
quasiparticle statistics of dopant-induced polaron carriers.
For the Holstein-$ tJ $ and Holstein-Hubbard models 
on the 8-, 10-, and 16-site clusters, the $d_{x^2-y^2}$-wave 
state is stable for two polarons. The second neighbor hopping 
$H_{t'}$ favors the $d_{x^2-y^2}$-wave pair for $t'>0$, while 
the long-range Coulomb repulsion $H_{\rm lc}$ favor the 
$ p_{x(y)} $-wave pair.

The strong on-site Hubbard-$U$ Coulomb repulsion plays a crucial
role in the formation of these pairing states. 
By keeping the electrons spatially
separated and preventing on-site bi-polaron formation the Hubbard-$U$
interaction acts, effectively, to greatly enhance the
polaron tunneling bandwidths and, hence, their mobility,
in the nearly $1\over2$-filled regime.

For a hypothetical superconducting polaron pair condensate, 
our results imply qualitative doping dependences
of the isotope effect, $ T_c $ and pseudo-gap
which are similar to those observed in the cuprates.
Potential limitations of the present polaron model, 
arising from the short-range nature of the assumed EP coupling, 
have been pointed out. Further studies to
include longer-range EP couplings, in combination with 
extended Coulomb interactions, have been outlined.

\acknowledgments

One of us (HBS) would like to thank D. Emin, J.P. Franck,
K. Levin and M. Norman for helpful discussions.
This work was supported by Grant No. DMR-9215123 
from the National Science Foundation 
and a Grant-in-Aid for Scientific Research on Priority Area 
``Nanoscale Magnetism and Transport'' 
from the Ministry of Education, Science, Sports and Culture, Japan.
Computing support from UCNS at the University of Georgia is 
gratefully acknowledged.

%
%
%%%%%%%%%%%%%%%%%%%%%%%%%%%%%%%%%%%%%%%%%%%%%%%%%%%%%%%%%%%%%%%%%
%
% FIGURE CAPTIONS
%
%%%%%%%%%%%%%%%%%%%%%%%%%%%%%%%%%%%%%%%%%%%%%%%%%%%%%%%%%%%%%%%%%
%
%

\begin{figure}\caption{
0th and 1st order contributions $W_0$ and $W_1$
to the effective adiabatic lattice potential in the
2-site Holstein model with 1 electron. 
In (a), 
the electronic groundstate energy $W_0/E_{\rm P}$ 
and the 1st excited state energy $W_0^{(1)}/E_{\rm P}$ of 
the electronic Hamiltonian $H_0(u)$ are shown
as functions of $u_-/u_{\rm P}$, 
at $u_+=0$ for 
$t/E_{\rm P}=0$, $0.1$, $0.3$, $0.5$ and $0.7$.
In (b), 
$W_1\times E_{\rm P}/\Omega^2$ vs. $u_-/u_{\rm P}$
is shown at $u_+=0$ for
$t/E_{\rm P}=0.1$, $0.3$, $0.5$ and $0.7$.
}\label{fig:double_well}\end{figure}%01

\begin{figure}\caption{
Exact and adiabatic results for the tunneling
energy splitting, $2t_P$, 
between the groundstate and the 1st excited electron-phonon
eigenstate, plotted as a function of $E_{\rm P}/\Omega$,
in the 2-site Holstein model with 1 electron,
for $t=1$ and several values of $E_{\rm P}$, as indicated.
}\label{fig:tunnel2site}\end{figure}%02

\begin{figure}\caption{
(a) One-polaron and (b) two-polaron closed tunneling paths and 
their Berry phase factors.
Black circles indicate the polaron locations for the initial 
$u$-configuration of the path.
The numbers on the two-polaron exchange paths in (b) indicate 
the order of the single-polaron tunneling steps.
}\label{fig:closed_path}\end{figure}%03

\begin{figure}\caption{
Important single-polaron tunneling processes with matrix elements 
$t_P^{(\nu)}$ to the $ \nu $-th neighbor sites for (a) P=1 and 
(b) P=2 polaron states on the 2D square lattice. 
}\label{fig:hopping}\end{figure}%04

\begin{figure}\caption{
(a) Berry phase contributions from single-polaron 1st-neighbor 
processes in the vicinity of a second, static polaron (black circle).
Full and dashed bonds indicate ($+1$) and ($-1$) Berry phase 
contributions, respectively.
(b) Internal parity of 2nd- and 3rd-neighbor polaron pairs 
is odd under reflection along the dashed line.
}\label{fig:parity}\end{figure}%05

\begin{figure}\caption{
(a) One-polaron and (b) two-polaron open tunneling paths and 
their Berry phase factors.
Black circles indicate the polaron locations for the initial 
$u$-configuration of the path.
$T$($ a $,$ b $) denotes the translation by vector ($ a $,$ b $), 
$R$($ \phi $) the rotation by angle $ \phi $.
}\label{fig:open_path}\end{figure}%06

\begin{figure}\caption{
Total momentum $(p_x, p_y)$ of the 
(a) P=1 and (b) P=2 polaron ground state of $H_{\rm P}$ 
in a $ t_P^{(3)} $-vs.-$ t_P^{(2)} $ phase diagram.
Also shown in (b) are the internal symmetries of 
the respective two-polaron ground states.
}\label{fig:phase}\end{figure}%07

\begin{figure}\caption{
(a) Logarithm of the effective polaron hopping amplitudes 
$t_P^{(\nu)}$ for $P$=1 and 2 holes and $ \nu $=2nd- and 
3rd-neighbor processes vs. inverse phonon energy $ 1/\Omega $ 
(with $t_P^{(2)}=t_P^{(3)}$ due to accidental cluster symmetries).
(b) Effective polaron nearest-neighbor attraction $V_{\rm P}$ 
and two-polaron binding energy $ \Delta $ vs. $ 1/\Omega $.
All results are for $t\equiv1$, $J=0.5t$, with 
$E_{\rm P}\equiv C^2/K=2.5t$ and $4.0t$, 
on an $N$=8 lattice with periodic boundary conditions.
}\label{fig:parameter}\end{figure}%08

\begin{figure}\caption{
(a) Logarithm of the effective polaron hopping amplitudes 
$t_P^{(\nu)}$ for $P$=1 and 2 holes and $ \nu $=2nd- and 
3rd-neighbor processes vs. inverse phonon energy $ 1/\Omega $ 
(with $t_P^{(2)}=t_P^{(3)}$ due to accidental cluster symmetries).
(b) Effective polaron nearest-neighbor attraction $V_{\rm P}$ 
and two-polaron binding energy $ \Delta $ vs. $ 1/\Omega $.
All results are for $t\equiv1$, $U=8t$, with 
$E_{\rm P}=2.5t$ and $4.0t$, 
on an $N$=8 lattice with periodic boundary conditions.
}\label{fig:Hubbard}\end{figure}%09

\begin{figure}\caption{
Logarithm of the effective polaron hopping amplitudes 
$t_2^{(\nu)}$ for $ \nu $=2nd- and 3rd-neighbor processes vs. 
inverse phonon energy $ 1/\Omega $, with and without 
inclusion of next-nearest-neighbor hopping $ t' $=$+0.2t$, 
for $t\equiv1$, $J=0.5t$, 
(a) $E_{\rm P}=2.5t$, and (b) $E_{\rm P}=4.0t$, 
on an $N$=8 lattice with periodic boundary conditions.
}\label{fig:2nd_n_hopping_plus}\end{figure}%10

\begin{figure}\caption{
Logarithm of the effective polaron hopping amplitudes 
$t_2^{(\nu)}$ for $ \nu $=2nd- and 3rd-neighbor processes vs. 
inverse phonon energy $ 1/\Omega $, with and without 
inclusion of next-nearest-neighbor hopping $ t' $=$-0.2t$, 
for $t\equiv1$, $J=0.5t$, 
(a) $E_{\rm P}=2.5t$, and (b) $E_{\rm P}=4.0t$, 
on an $N$=8 lattice with periodic boundary conditions.
}\label{fig:2nd_n_hopping_minus}\end{figure}%11

\begin{figure}\caption{
Logarithm of the effective polaron hopping amplitudes 
$t_2^{(\nu)}$ for $ \nu $=2nd- and 3rd-neighbor processes vs. 
inverse phonon energy $ 1/\Omega $, with and without 
inclusion of long-range repulsion $ V_{\rm C} $=$1.0t$, 
for $t\equiv1$, $J=0.5t$, 
(a) $E_{\rm P}=2.5t$, and (b) $E_{\rm P}=4.0t$, 
on an $N$=8 lattice with periodic boundary conditions.
}\label{fig:long_range_Coulomb}\end{figure}%12


\begin{references}
\bibitem{YZS1}
In Ref.~\onlinecite{YZS}, the momenta of the $ p $-wave 
ground states were given incorrectly.
\end{references}
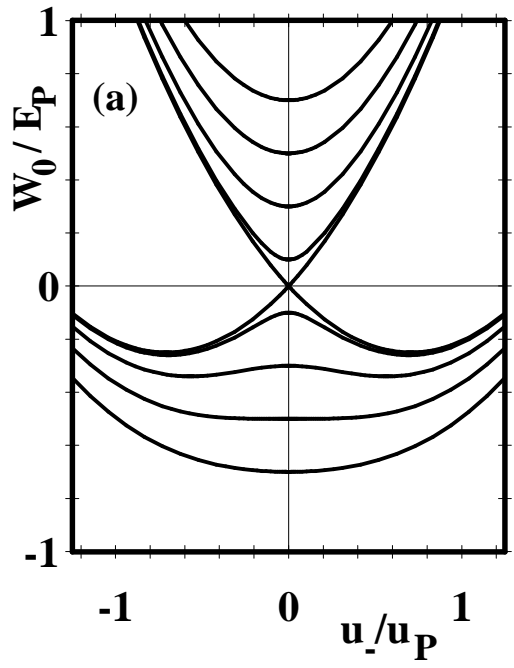
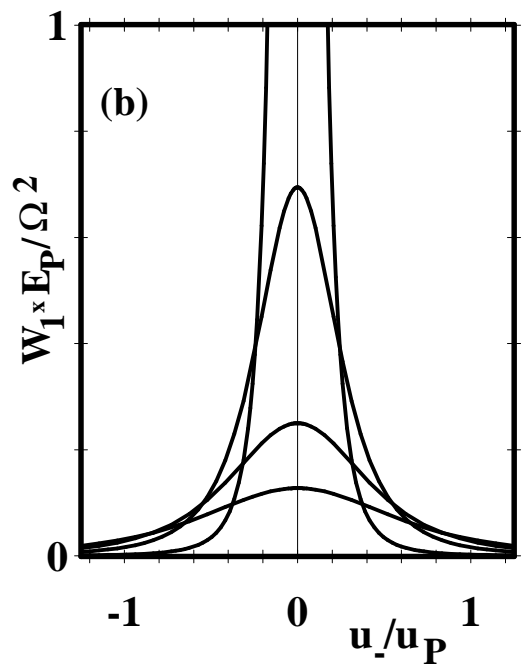
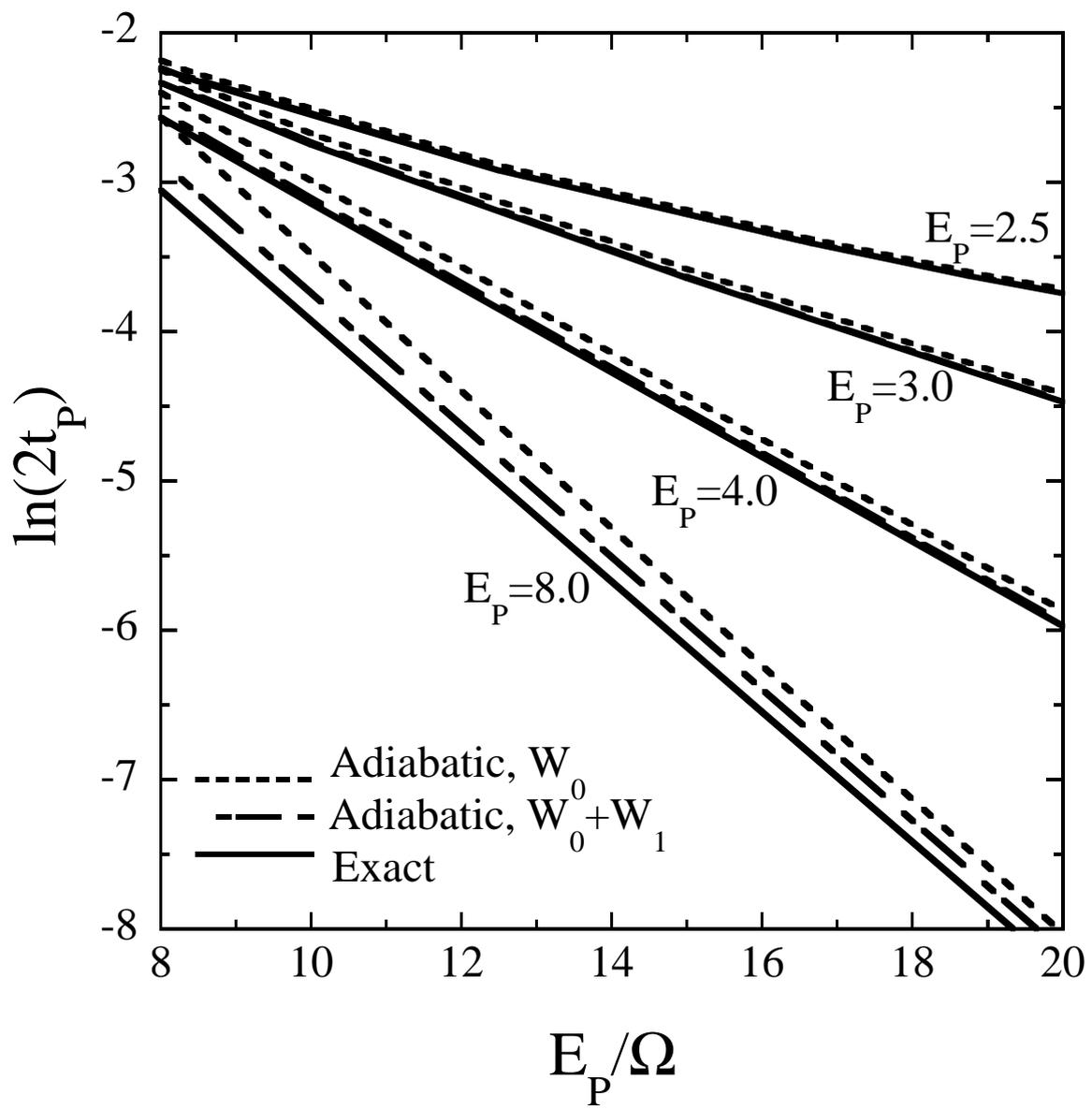
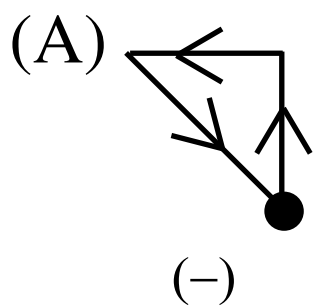
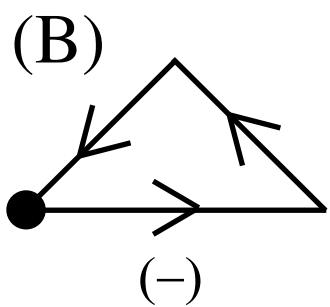
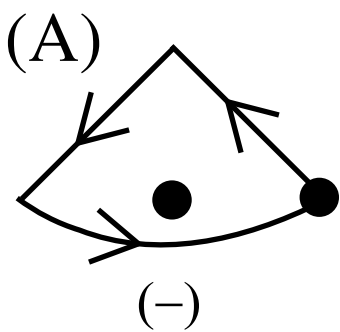
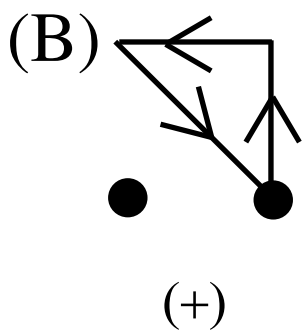
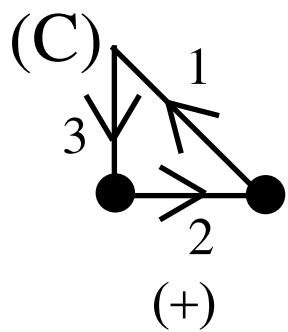
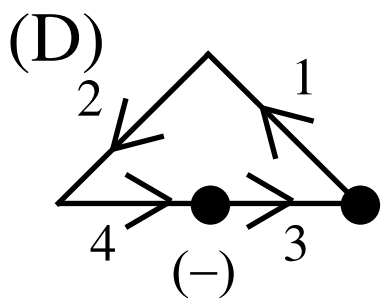
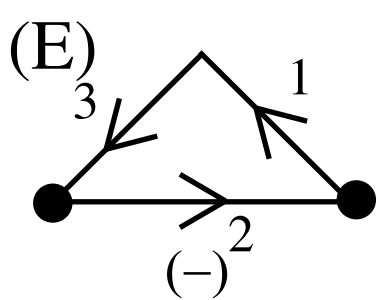
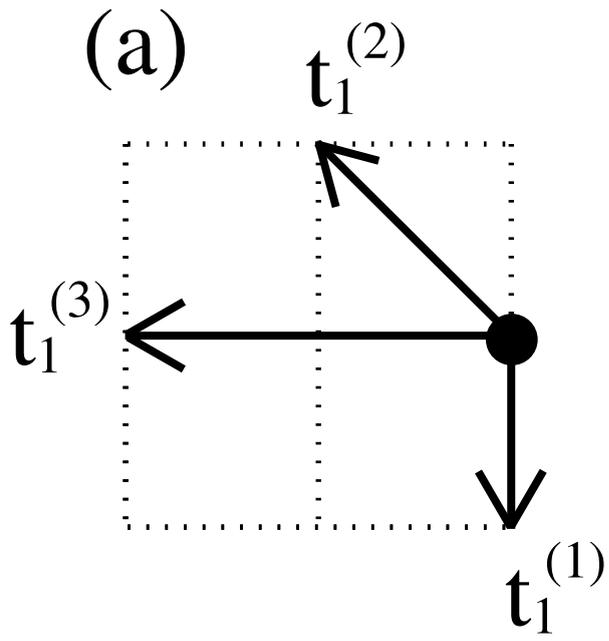
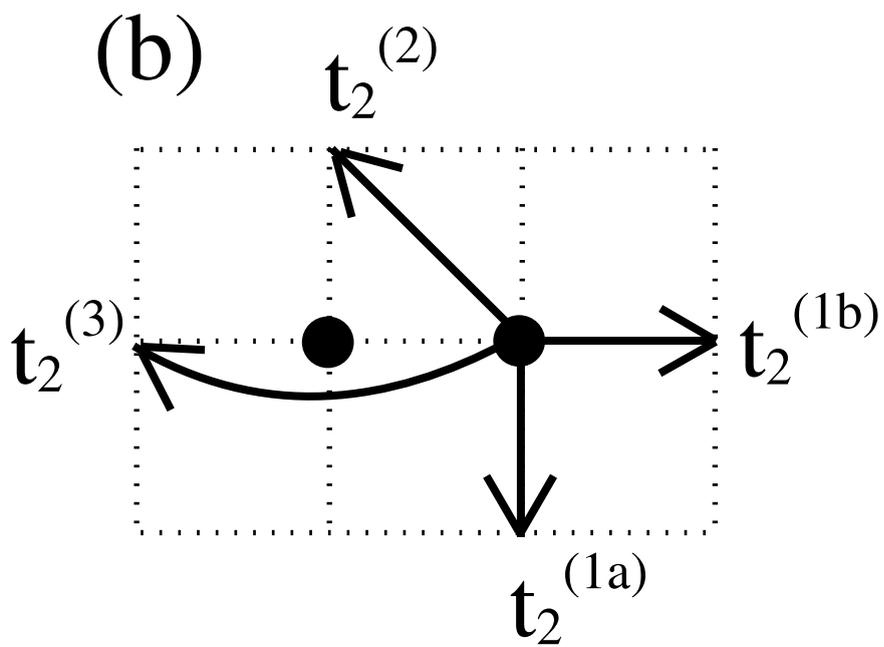
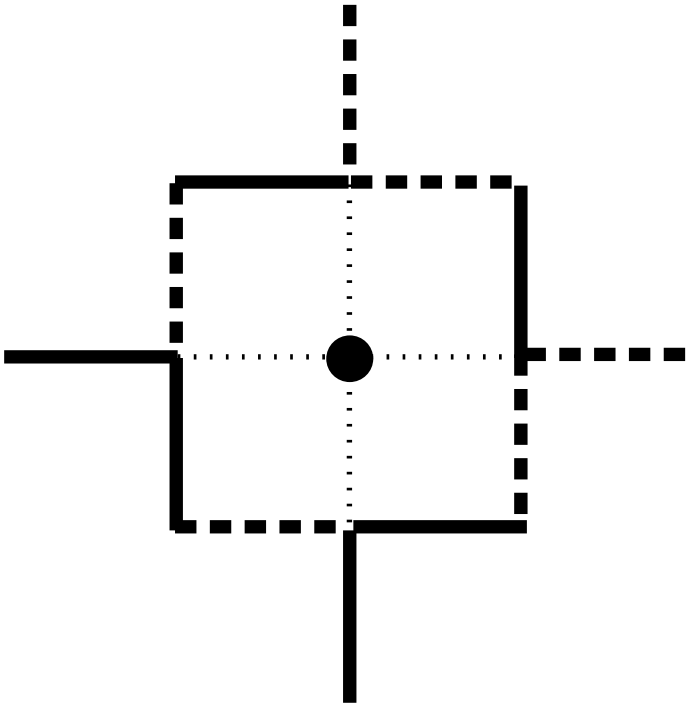
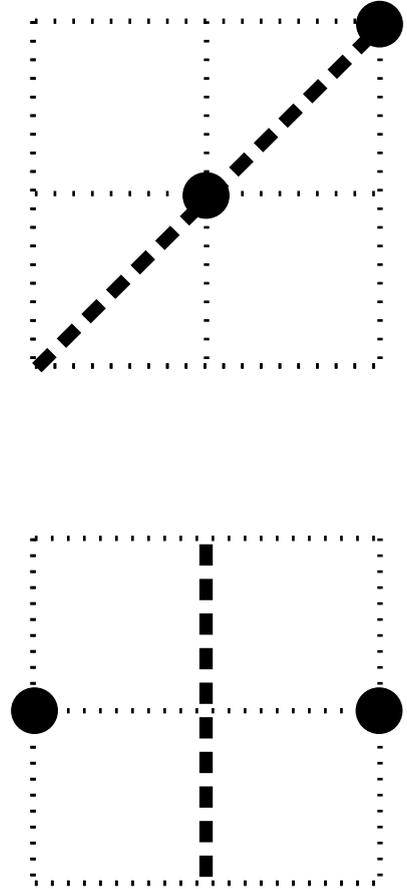
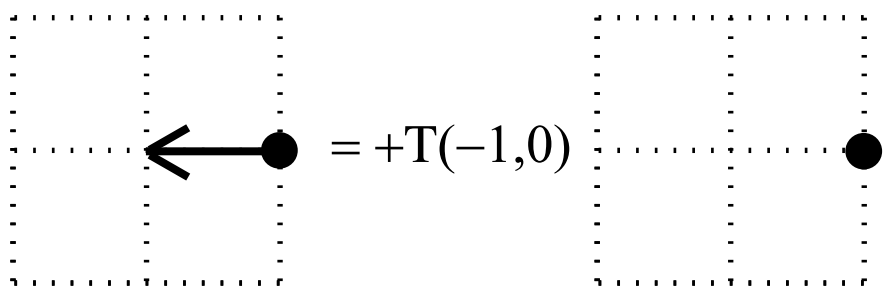
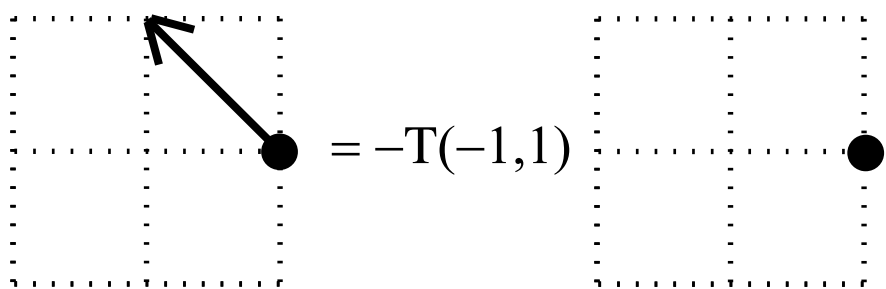
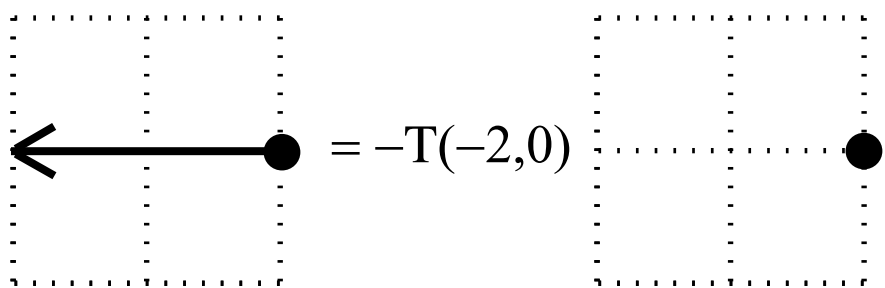
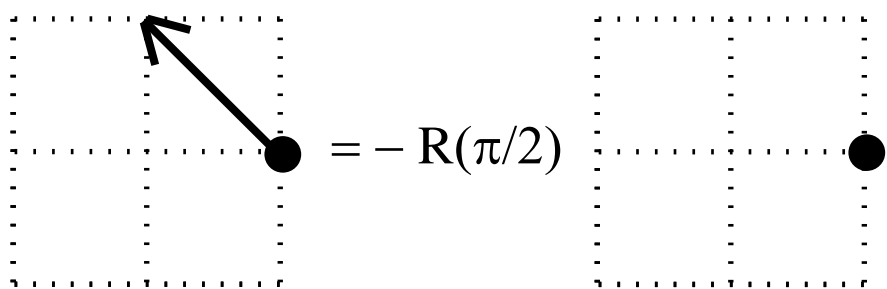
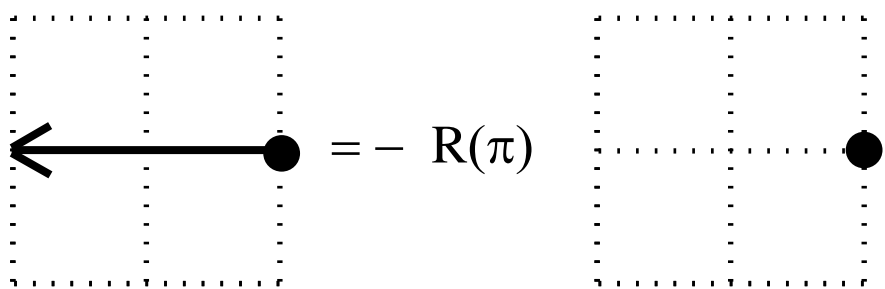
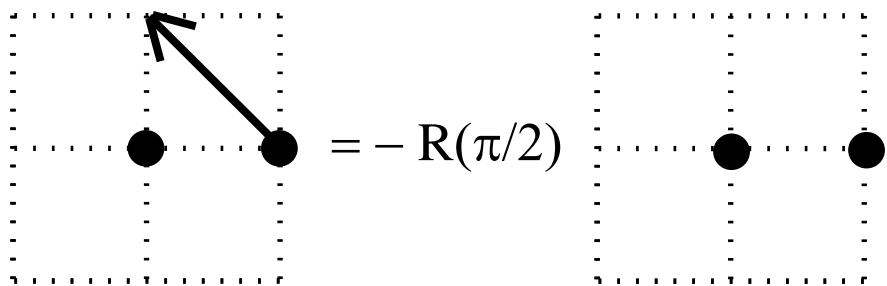
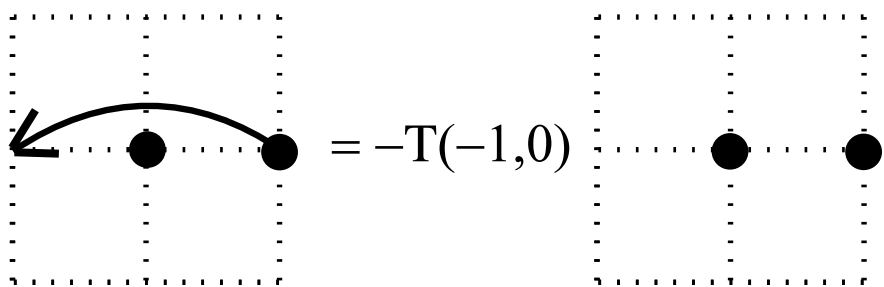
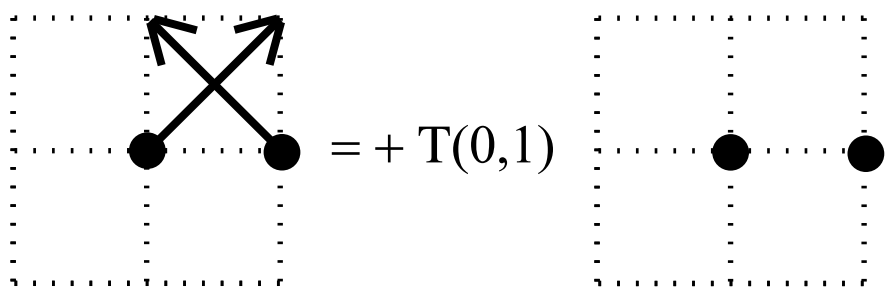
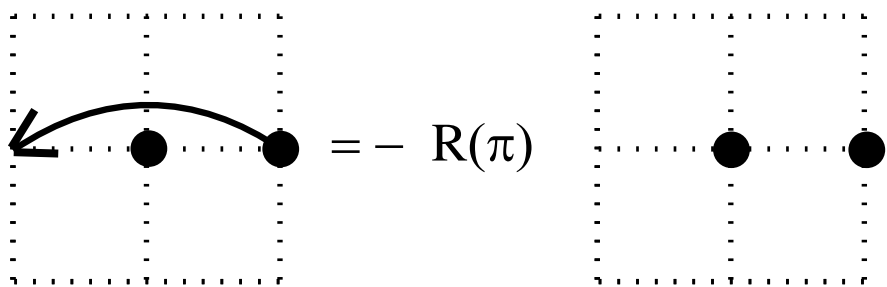
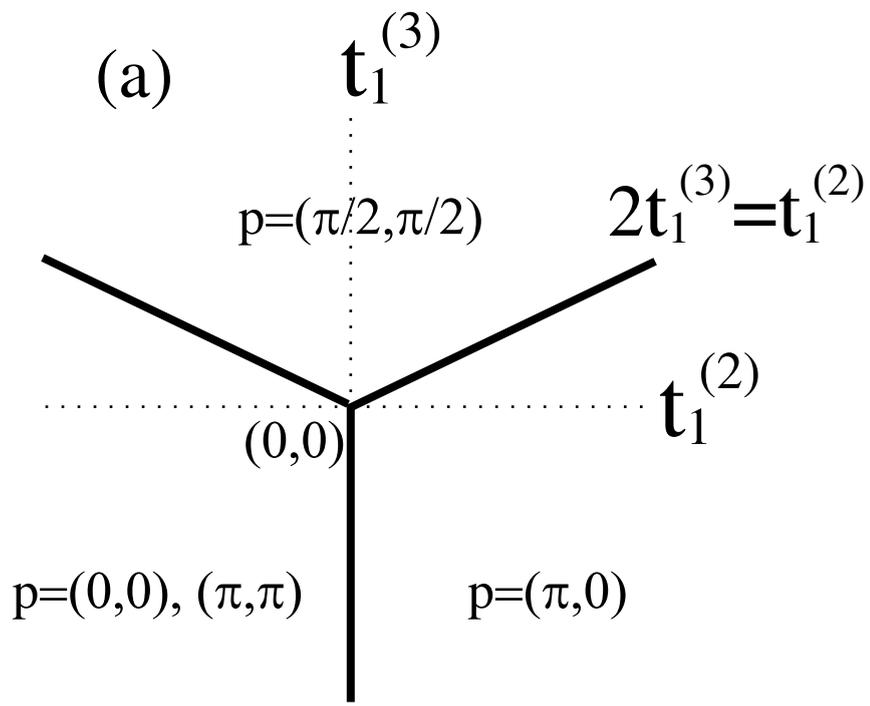
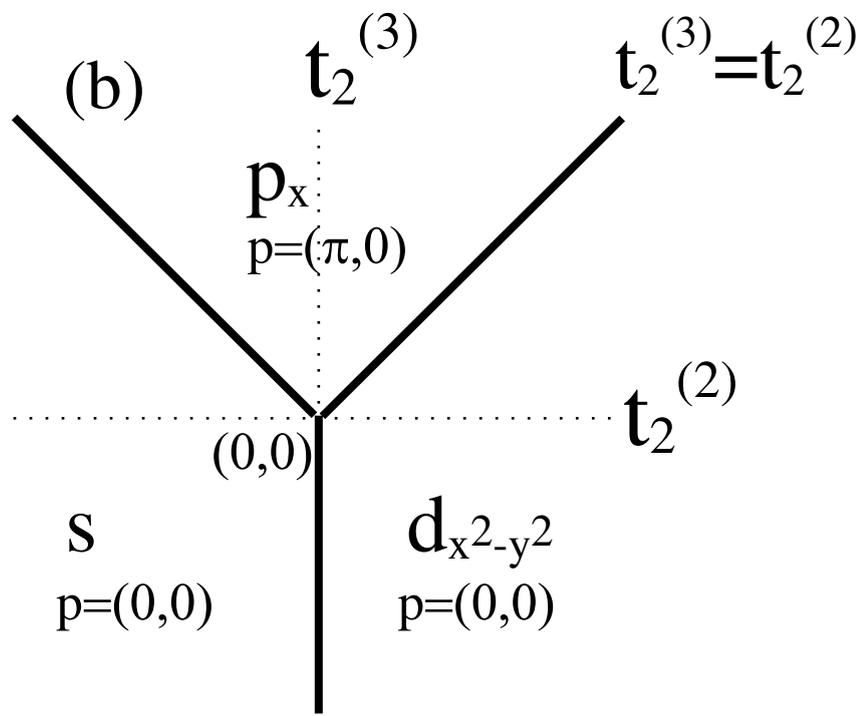
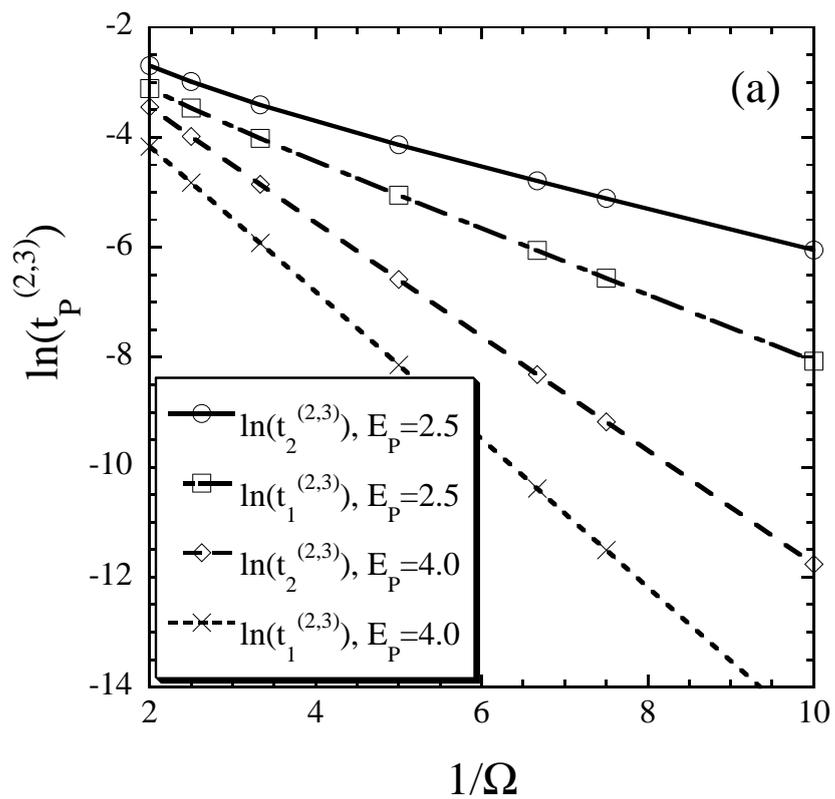
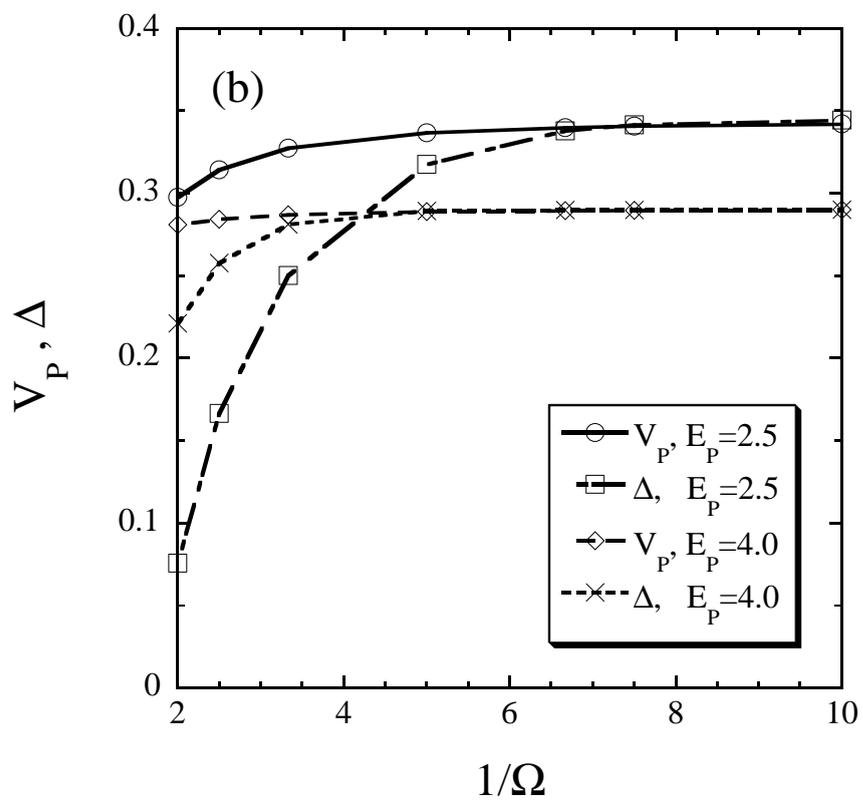
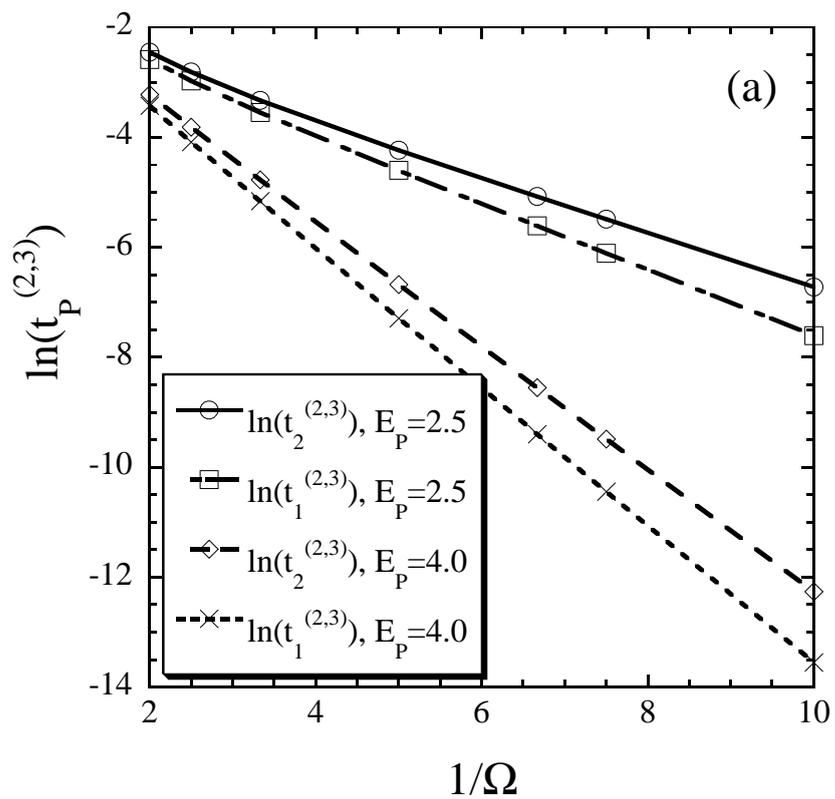
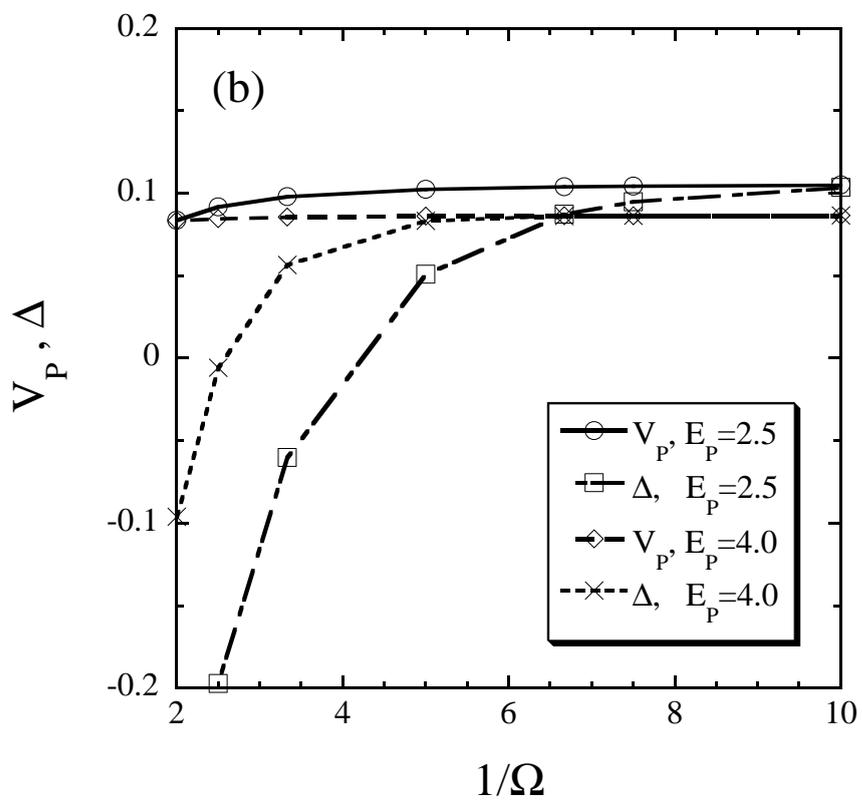
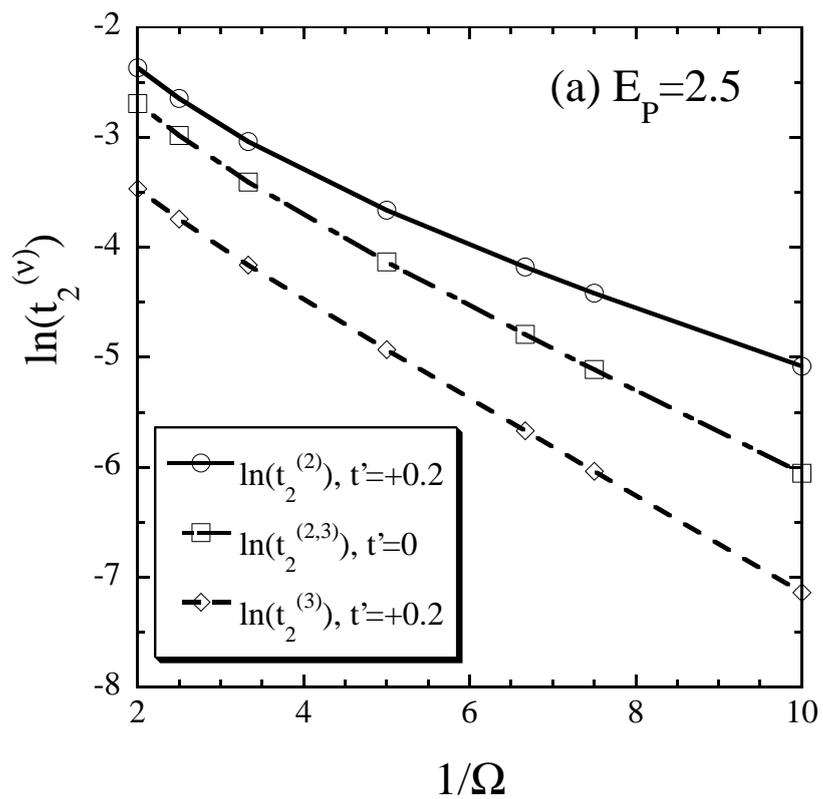
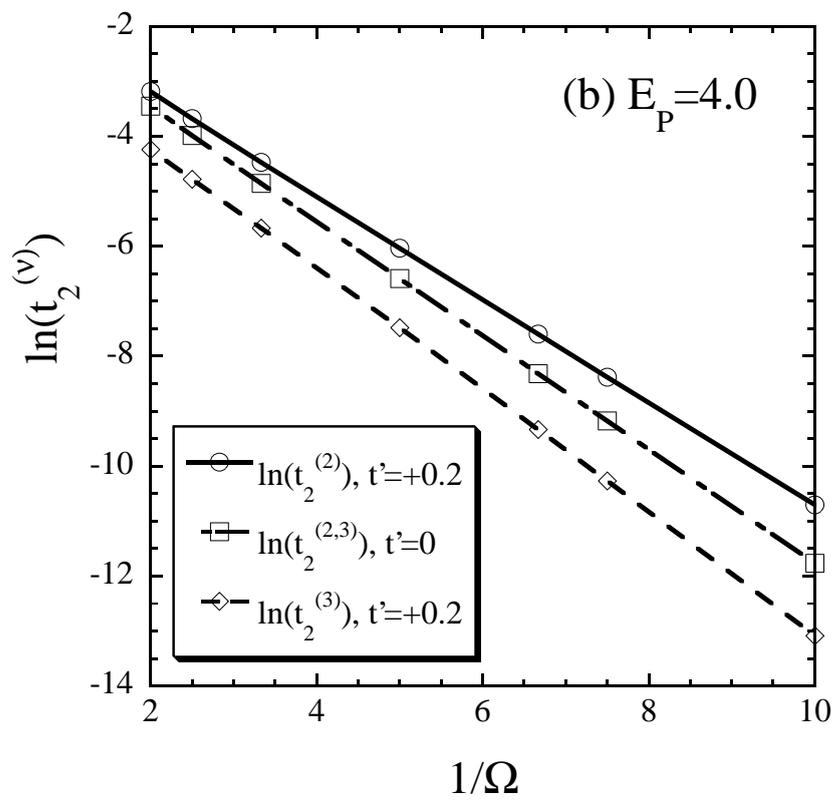
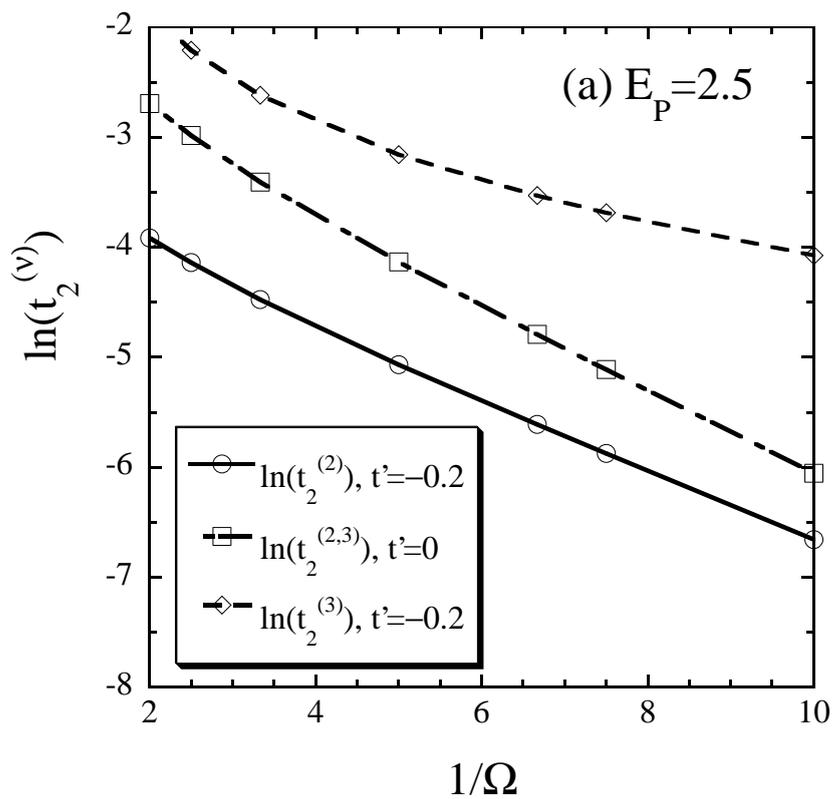
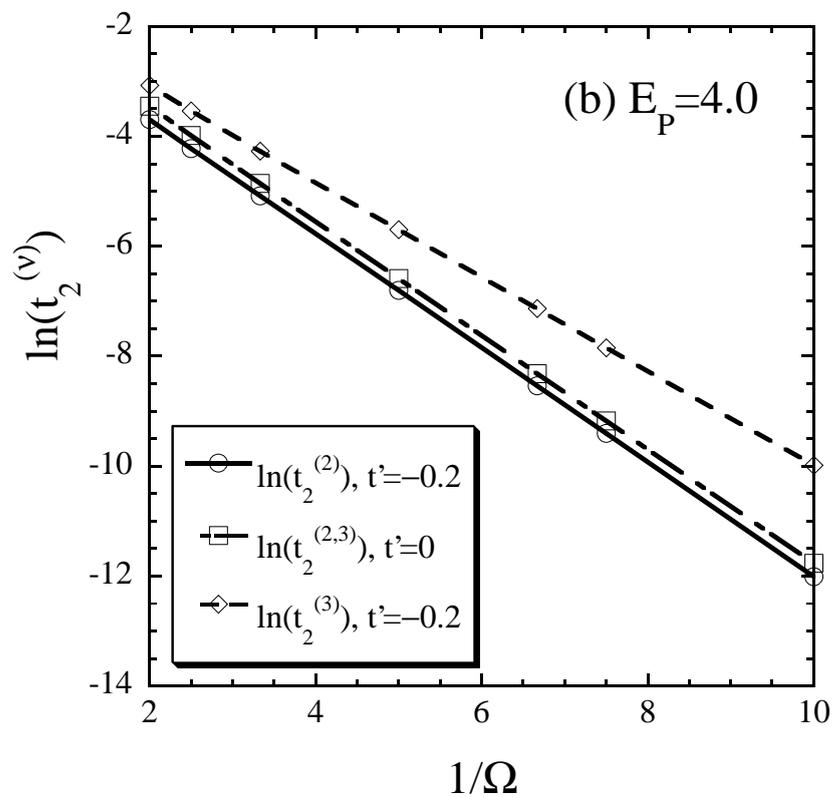
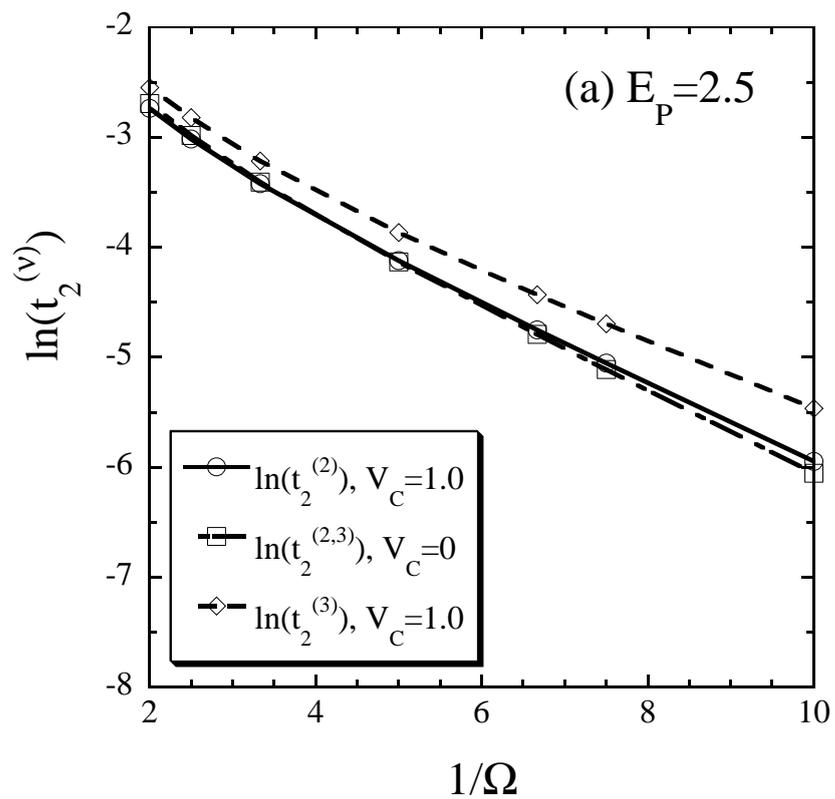
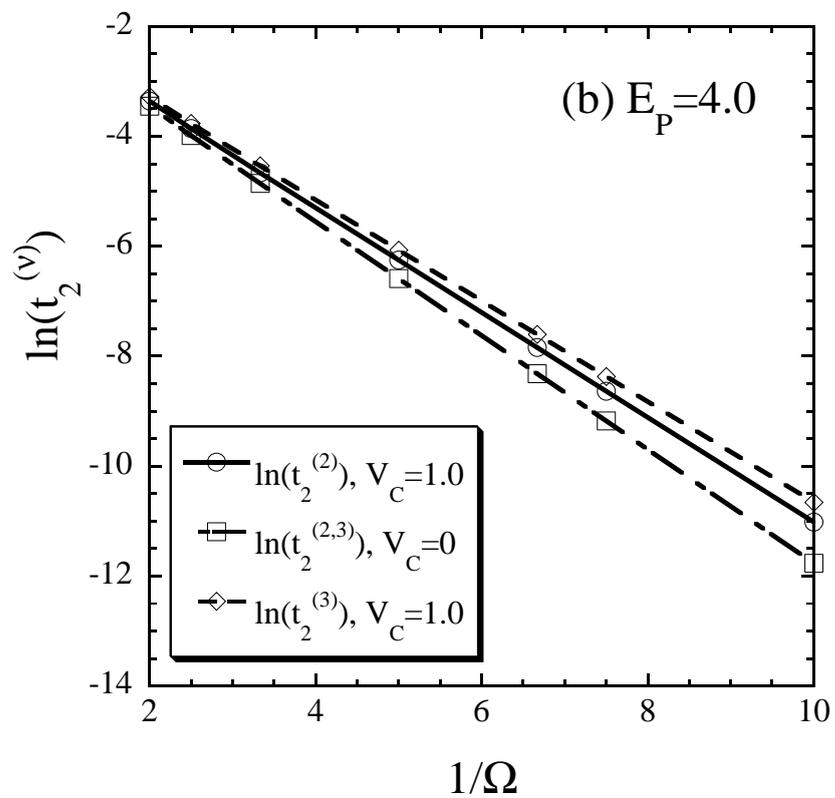
\end{document}